\documentclass[aps,pra,superscriptaddress,amsmath,amssymb,preprintnumbers,floatfix,showpacs,12pt]{revtex4}
\usepackage{amssymb}
\usepackage{epsfig}
\usepackage{subfigure}
\usepackage{graphicx}
\begin{document}
\title{ Evolution  of  cat states in a dissipative parametric amplifier:
decoherence and entanglement
 }

\author{Faisal A. A. El-Orany }
\email{el_orany@hotmail.com, Tel:006-166206010, Fax:
0060386579404}\affiliation{ Department of Mathematics and Computer
Science, Faculty of Natural Science, Suez Canal University,
Ismailia, Egypt; Department of Optics, Palack\'y University,
17.~listopadu~50, 772 07~Olomouc, Czech Republic}
\author{
J.Pe\v{r}ina\footnote{Joint Laboratory of Optics of Palack\'y
University and Institute of Physics,  Academy of Sciences of the
Czech Republic, 17.~listopadu~50, 772 07~Olomouc, Czech
Republic.}, V. Pe\v{r}inov\'{a}}

\affiliation{ Department of Optics, Palack\'y University,
17.~listopadu~50, 772 07~Olomouc, Czech Republic}

\author{ and M. Sebawe Abdalla}

\affiliation{Mathematics Department, College of Science, King Saud
University, P.O. Box 2455, Riyadh 11451, Saudi Arabia}

\begin{abstract}
The evolution of the Schr\"{o}dinger-cat states in  a dissipative
parametric amplifier is examined. The main tool  in the analysis
is the normally ordered characteristic function. Squeezing,
photon-number distribution and reduced factorial moments
 are discussed for the single- and compound-mode cases.
Also the single-mode Wigner function  is  demonstrated. In
addition to the decoherence resulting from the interaction  with
the environment (damped case) there are two  sources which can
cause such decoherence in the system even if it is completely
isolated: these are the decay of the pump and
 the relative phases of the initial cat states.
Furthermore, for the damped case there are  two regimes, which are
 underdamped and overdamped.
In the first (second) regime the signal mode or the idler mode
``collapses" to a statistical mixture (thermal field).

\end{abstract}
 \maketitle


\section{Introduction}
 The linear superposition principle is at the heart of quantum mechanics
 since, using it, one can control the properties of the single states making
them more or less pronounced.
The most significant
example reflecting the power of such a principle are the
Schr\"{o}dinger-cat
states \cite{sch1}, which exhibit various nonclassical effects
such as squeezing, sub-Poissonian statistics and oscillations in photon-number
distribution \cite{yur, {yur1},{micro1}}, even
if the original states are close to the classical ones
\cite{buz1}.
Such  type of states (of a $j$th mode for convenience)
can be represented as

\begin{equation}
|\alpha\rangle_{\phi_{j}}=N_{j} [|\alpha_{j}\rangle+\exp(i\phi_{j})
|-\alpha_{j}\rangle],\label{1}
\end{equation}
where $|\alpha_{j}\rangle$ is a coherent state of the $j$th mode
with complex amplitudes $\alpha_{j}$,
$\phi_{j}$ is a relative phase and  $N_{j}$ is the normalization constant
having the form

\begin{equation}
N^{2}_{j}=\frac{1}{2[1+ \exp(-2|\alpha_{j}|^{2})\cos \phi_{j}]}.\label{2}
\end{equation}
Specifically there are three  values of  $\phi_{j}$, namely,
$0,\pi$ and $\pi/2$, for which (\ref{1}) reduces to  even coherent (ECS),
odd coherent
(OCS) and Yurke-Stoler (YSS) states, respectively. These three states
will be frequently used in this paper.

There are several proposals for generating  states (\ref{1})
in various nonlinear processes, e.g. see \cite{micro1,{gen},{back},{micro2},
{scha},{zheng1}}.
Further, the evolution of  cat states in several quantum systems
has been intensively  studied
\cite{buz2}--\cite{filip}. For convenience of the present work we refer to
the interaction of these states with
environment \cite{buz2,{buz3},{buz4},{buz5},{filip}}, where
  the common goal of such studies is the comprehension and
description of the decoherence processes
(decoherence is the rapid transformation of a pure
linear superposition state into the corresponding statistical mixture
state)
in the system under
observation, and the energy loss due to the interaction with the
reservoir.
Most of these papers adopt the Heisenberg--Langevin approach or the Markov
master equation approach. Although the non-Markovian dynamics is important
in  quantum physics (theory), a unified and compact treatment of the
non-Markovian reservoirs dates as late from \cite{gisin}, where also the
decay and revival of the Schr\"{o}dinger-cat states are characterized.

On the other hand,  parametric amplifier (nondegenerate) takes a considerable
interest in quantum optics since it can
 perfectly generate two-mode squeezing.
Recently, this device has been supported with the
fast progress of new nonlinear crystals and  improved laser sources, especially
femtosecond lasers \cite{laser} and it
  has been employed in experiments,
e.g. in the interference experiments \cite{kwia,{zou1}}.
An investigation of the statistical properties
 of the  parametric amplifier
with losses \cite{in5} and without losses \cite{in2,{in3},{in4}}
for compound-mode case have  been performed using different
techniques when the modes are initially prepared  in the coherent states.
The anticorrelation in this model is an interesting effect
\cite{in4}, where also under certain conditions the variance of
the photon number is less than the average photon number and the
photocounting distribution becomes narrower than the corresponding
Poisson distribution for a coherent state with the same mean photon
number. Moreover, the sum photon-number distribution can exhibit
collapses and revivals \cite{faisal} (and references therein) similar to
those typical in the Jaynes-Cummings model (JCM) \cite{cum}; the former
is in the photon number domain rather than the time domain. So the
question we would like  to address in this paper is the following: what
would be the influence of the interference in phase space on these
phenomena? In other words, we study the evolution of the cat states
(\ref{1}) through the parametric amplifier and take into account the
interaction  with environment.  The system alone may be described by
tracing a more complete description
over the environment. Actually, the interaction considered here
is quite different from that  considered in  \cite{buz2}--\cite{buz6}
because there are two operations controlling its  behaviour, which are
the interference in phase space and the entanglement (correlation)
between the signal and idler modes as well as between the signal-idler
system and the environmental oscillators (reservoirs) to which the
system may be connected. Such entanglements  lead to the increase of the
marginal entropies of the subsystems (as is known, if the density
operator describes  pure states, then the entropy ($S$) is zero,
otherwise $S\neq 0$) \cite{pho}). For  the present system we treat the
single- and compound-mode cases. The main tools in our treatment are the
normally ordered characteristic functions. The Hamiltonian which governs
the interaction is \cite{in9,{in10}}
\begin{eqnarray}
\begin{array}{lr}
\hat{H}=\hbar \sum\limits_{j=1}^{2}\omega_{j}\hat{a}^{\dagger}_{j}\hat{a}_{j}
-\hbar g[\hat{a}_{1}\hat{a}_{2}\exp(i\omega t-i\phi) +{\rm H.c.}]\\
\\
+\hbar \sum\limits_{j=1}^{2}\sum\limits_{l=1}^{\infty}[\varphi_{jl}\hat{b}_{jl}^{\dagger}
\hat{b}_{jl}+k_{jl}\hat{b}_{jl}\hat{a}^{\dagger}_{j}
 +k^{*}_{jl}\hat{b}_{jl}^{\dagger}\hat{a}_{j}],
\label{3}
\end{array}
\end{eqnarray}
where $\hat{a}_{j}, (\hat{a}^{\dagger}_{j}),\quad j=1,2$,
are the annihilation (creation) operators
assigned  to the signal and idler modes, respectively;
$\omega_{j}$ are the natural  frequencies of oscillations
of the uncoupled modes with $\omega =\omega_{1}+\omega_{2},
\omega$  being a pump frequency (under the assumption of
 resonance frequency); $\phi$ is the initial phase of
the pump and {\rm H.c.} means the Hermitian conjugate term to the
previous one; $g$ is the real coupling
constant including the amplitude of the pump (gain coefficient);
$\hat{b}_{jl}$ and $\hat{b}_{jl}^{\dagger}$ are the boson annihilation and
creation
operators of the reservoir oscillators, respectively, with the
frequencies
$\varphi_{jl}$, and $k_{jl}$ are the coupling constants of the
system-reservoirs interaction.  Hamiltonian (\ref{3}) represents three
wave mixing in a nonlinear crystal described by the second-order susceptibility
$\chi^{(2)}$. Specifically, a pump beam with frequency $\omega$
travelling through the crystal  creates photons of the signal and idler modes
with different frequencies $\omega_{j}$, $j=1,2$, such that
$\omega=\omega_{1}+\omega_{2}$.  The whole system is considered to be
interacting with its surroundings (e.g., via imperfect cavity mirrors),
which are modelled as reservoirs.

For completeness, the well-known solution of the Heisenberg--Langevin equations
for the Hamiltonian (\ref{3}) is  \cite{in9,{in10}}

\begin{eqnarray}
\begin{array}{lr}
\hat{A}_{1}(t)
= f_{1}(t)\hat{a}_{1}(0)+f_{2}(t)\hat{a}^{\dagger}_{2}(0)
+\sum\limits_{l}^{\infty}[\hat{b}_{1l}(0)\Gamma_{1l}(t)
+\hat{b}_{2l}^{\dagger}(0)\Gamma^{'}_{1l}(t)],\\
\\
\hat{A}_{2}(t)
 = f_{3}(t)\hat{a}_{2}(0)+f_{2}(t)\hat{a}^{\dagger}_{1}(0)
+\sum\limits_{l}^{\infty}[\hat{b}_{1l}^{\dagger}(0)\Gamma_{2l}(t)
+\hat{b}_{2l}(0)\Gamma^{'}_{2l}(t)],
\label{4}
\end{array}
\end{eqnarray}
where $\hat{A}_{j}(t)=\hat{a}_{j}(t)\exp(i\omega_{j}t)$;
 $\hat{a}_{j}(0)$ and $ \hat{b}_{jl}(0), \quad j=1,2,$
 are the initial operators  of the modes and reservoirs, respectively.
The explicit forms for the dynamical coefficients
$f_{j}(t),\Gamma_{jl}(t)$ and $\Gamma^{'}_{jl}(t)$ are given in the appendix A.
Further, it is worth mentioning that the Langevin forces
have the forms

\begin{equation}
\hat{L}_{j}(t)=-i\sum\limits_{l}k_{jl}\hat{b}_{jl}(0)\exp(-i\varphi_{jl}t), \quad j=1,2.
\label{5}
\end{equation}
These forces satisfy the following commutation rule
\begin{equation}
[\hat{L}_{j}(t),\hat{L}^{\dagger}_{j'}(t')]=\gamma_{j}
\delta_{jj^{'}}\delta(t-t')\hat{1},
\label{6}
\end{equation}
where $\gamma_{j}$ is the cavity decay rate of the $j$th mode
 (for more details about
the properties of the reservoir oscillators, see \cite{in3,{in11}}).
 Finally it is worth mentioning that including  lossy mechanism,
the difference mean-photon number between the signal and idler modes
in the system becomes nonconservative.

This paper is organized as follows: In section 2 the basic
equations and
relations, such as two-mode normally ordered characteristic function,
quadrature squeezing, Wigner function,  reduced factorial moments
and photon-number distribution are given when the modes are
initially prepared in the cat states (\ref{1}).
In sections 3 and 4 discussion of the results of the single-mode
and compound-mode cases are
performed, respectively, and the conclusions are summarized in section
5.

\section{Basic relations and equations}

In this section we give the main relations which will be
used to investigate the properties of the system under consideration. Actually,
the calculations are  lengthy, but straightforward, and
for this reason, we give briefly general calculations in such a way that
particular results can be obtained using  suitable choice of  parameters.
We start by discussing the initial density matrix
of the system.
We assume that the system and  reservoirs are initially independent and
noninteracting before switching on  the interaction, and the
interaction between them starts at $t=0$.
This means that the density operator could be written initially as a direct
product \cite{wolf}:
\begin{equation}
\hat{\rho}(0)=\hat{\rho}_{\rm f}(0)\otimes\hat{\rho}_{\rm r}(0),
\label{7}
\end{equation}
where $\hat{\rho}_{\rm r}(0) $ is the density matrix of the
system of damping oscillators,
which we assume to have flat (i.e. constant function of
 frequency) and broad-band  reservoir spectra, so that
the mean number of reservoir quanta (phonons) in the mode $l$
is $ \langle \hat{b}^{\dagger}_{jl}(0)\hat{b}_{jl}(0)\rangle=
\langle \hat{n}_{jd}\rangle$
independently of $l$, where the subscript $d$ denotes broad-band reservoir.
 Also we assume that
the reservoirs form  chaotic systems with  mean photon
numbers $\langle \hat{n}_{jd}\rangle=
[\exp(\frac{\hbar\varphi_{j}}{T K_{B}})-1]^{-1},$
where the reservoirs-oscillators are at the temperature $T$, with the
frequency $\varphi_{j}$,
$K_{B}$ is  Boltzmann's constant
 and $\hbar$  Planck's constant divided by $2\pi$.
We proceed assuming that
 $\hat{\rho}_{\rm f}(0)$ is the field density operator, which in our case,
where the  modes are assumed to be initially in
superposition states (\ref{1}),
has the form
\begin{eqnarray}
\begin{array}{lr}
\hat{\rho}_{\rm f}(0)=
|\alpha\rangle_{\phi_{1}}|\alpha\rangle_{\phi_{2}}  \mbox{$_{\phi_{2}}$}\langle\alpha|
\mbox{$ _{\phi_{1}}$}\langle\alpha| \\
\\
=N^{2}_{1} N^{2}_{2}\sum\limits_{j',j=1}^{4}\exp(i\phi_{j'}+i\phi'_{j})
  |\alpha_{j'}\rangle |\alpha_{j}\rangle \langle\alpha'_{j}|
\langle\alpha'_{j'}| \\
  \\
  =N^{2}_{1} N^{2}_{2}[ \hat{\rho}_{\rm M}(0) +\hat{\rho}_{\rm SI}(0)
  +\hat{\rho}_{\rm AI}(0) ].    \label{8}
\end{array}
\end{eqnarray}
The concise form in the second line of (\ref{8}) means that the values of
$\alpha_{j},\phi_{j},\alpha'_{j}$ and $\phi'_{j}$ should be
specified to obtain
the exact form of the density operator in the first line.
Here $\hat{\rho}_{\rm M}(0)$ denotes  the statistical mixture part of the
 density operator as
\begin{eqnarray}
\begin{array}{l}
\hat{\rho}_{\rm M}(0)=
  |\alpha_{1}\rangle |\alpha_{2}\rangle \langle\alpha_{2}| \langle\alpha_{1}|+
  |\alpha_{1}\rangle |-\alpha_{2}\rangle \langle-\alpha_{2}| \langle\alpha_{1}|\\
\\
+|-\alpha_{1}\rangle |\alpha_{2}\rangle \langle\alpha_{2}| \langle-\alpha_{1}|+
|-\alpha_{1}\rangle |-\alpha_{2}\rangle \langle -\alpha_{2}| \langle -\alpha_{1}|;
  \label{9}
\end{array}
\end{eqnarray}
$\hat{\rho}_{\rm SI}(0)$ denotes  symmetric interference part of the
density operator in which the two modes are in off-diagonal basis of
coherent states and can be written as
\begin{eqnarray}
\begin{array}{l}
\hat{\rho}_{\rm SI}(0)=
\Bigl\{ \exp[i(\phi_{1}+\phi_{2})]  |-\alpha_{1}\rangle
|-\alpha_{2}\rangle \langle\alpha_{2}| \langle\alpha_{1}|
\\\\
+
\exp[-i(\phi_{1}+\phi_{2})]  |\alpha_{1}\rangle |\alpha_{2}\rangle \langle-\alpha_{2}|
\langle-\alpha_{1}|\Bigr\}+ {\rm H.c.}
\end{array}
\label{10}
\end{eqnarray}
and $\hat{\rho}_{\rm AI}(0)$ denotes  asymmetric interference part of the
density operator in which  one of the modes is  in a diagonal coherent
states basis while  the other is in off-diagonal basis  or vice versa
and reads:
\begin{eqnarray}
\begin{array}{lr}
\hat{\rho}_{\rm AI}(0)= \Bigl\{
\exp(-i\phi_{1}) \Bigl[|\alpha_{1}\rangle |\alpha_{2}\rangle \langle \alpha_{2}|
\langle -\alpha_{1}| + |\alpha_{1}\rangle |-\alpha_{2}\rangle \langle
-\alpha_{2}| \langle -\alpha_{1}|\Bigr]
\\\\
+ \exp(-i\phi_{2}) \Bigl[
|\alpha_{1}\rangle |\alpha_{2}\rangle \langle -\alpha_{2}| \langle
\alpha_{1}|
+ |-\alpha_{1}\rangle |\alpha_{2}\rangle \langle -\alpha_{2}|
\langle -\alpha_{1}|\Bigr]\Bigr\} + {\rm H.c.}.
\end{array}
\label{10ad}
\end{eqnarray}

The quantum properties of the system can be traced
via  the evolution of the normally ordered characteristic function,
which for the two-mode case has the form
\begin{equation}
C_{\cal N}(\zeta_{1},\zeta_{2},t)={\rm Tr} \left\{
\hat{\rho}(0) \prod_{j=1}^{2}\exp [\zeta_{ j}
\hat{A}_{j}^{\dagger}(t)]\exp [-
\zeta_{j}^{*} \hat{A}_{j}(t)] \right\},  \label{11}
\end{equation}
where, for the system under consideration,  $\hat{A}_{j}(t)$
are given in (\ref{4}) and
$\hat{\rho}(0)$ is the initial  density operator of
the system (\ref{7}).
It is obvious  that the characteristic function here includes 16
elements \cite{luk}. We give only the calculations related to  the element
$\exp(i\phi_{1}+i\phi'_{2})
|\alpha_{1}\rangle |\alpha_{2}\rangle \langle\alpha'_{2}| \langle\alpha'_{1}|$
of the density matrix.
On substituting this element together with  (\ref{4})  into (\ref{11}) and
calculating the expectation values, we arrive at
\begin{eqnarray}
\begin{array}{lr}
I(\zeta_{1},\zeta_{2},t)=
\exp\left[
i\phi_{1}+i\phi'_{2}
-\frac{1}{2}\sum\limits_{j=1}^{2}(|\alpha_{j}|^{2}
+|\alpha'_{j}|^{2}-2\alpha^{*}_{j}\alpha'_{j})\right]
\\
\\
\times\exp \left\{
\zeta_{ 1}\zeta_{ 2}D(t)+ \zeta^{*}_{ 1}\zeta^{*}_{
2}D^{*}(t)
-[\sum\limits_{j=1}^{2}|\zeta_{ j}|^{2}B_{j{\cal N}}(t)]
+\sum\limits_{j=1}^{2}[\zeta_{ j}\bar{\alpha}_{j}(t) -\zeta^{*}_{
j}\bar{\alpha}'_{j}(t)]
 \right\},  \label{12}
\end{array}
\end{eqnarray}
the explicit forms of the quantities $B_{j{\cal N}}(t), D(t), D^{*}(t)$ and
$\bar{\alpha}_{j}(t)$ are given in the appendix B.
It is worth mentioning that
throughout  the calculation of (\ref{12}),
the standard commutator
of the Langevin forces (\ref{6}) and
the usual Wigner-Weisskopf approximation have been used, which leads to
a replacement of the coupling constants by the cavity decay rates
$\gamma_{j}\geq 0$.
More details about the calculation of (\ref{12})
can be found in \cite{in9,{in10}}.
 On the other hand, the characteristic
function  of the single-mode case
can be obtained from that of the two-mode case by simply setting
the parameter related to the absent mode
equal to zero, e.g. the characteristic
function  of the
signal mode (first-mode) can be obtained by setting $\zeta_{2}=0$ in (\ref{11}) and
(\ref{12}).

The  different moments of the bosonic operator
 system  can be determined
by differentiation of the normally ordered characteristic function
through the relation
\begin{equation}
\langle \prod _{j=1}^{2}\hat{A}_{j}^{\dagger m_{j}}(t)\hat{A}_{j}^{n_{j}}(t)
\rangle= \prod _{j=1}^{2}\frac{\partial^{m_{j}+n_{j}}}
                           {\partial \zeta _{j}^{m_{j}}
                            \partial (-\zeta
_{j}^{*})^{n_{j}}}C_{\cal N}
                             (\zeta_{1},\zeta_{2},t)
                             |_{\zeta_{1}=\zeta_{2}=0}.  \label{13}
\end{equation}

In order to investigate squeezing property  for the compound-mode case
 we can define the two quadrature operators
$\hat{X}(t)=\frac{1}{2}\sum\limits_{j=1}^{2}[\hat{A}_{j}(t)+\hat{A}_{j}^{\dagger}(t)], \quad
\hat{Y}(t)=\frac{1}{2i}\sum\limits_{j=1}^{2}[\hat{A}_{j}(t)-\hat{A}_{j}^{\dagger}(t)]$,
where
$[ \hat{X}(t),\hat{Y}(t)] =i\hat{1}$
and then the uncertainty relation reads
$\langle (\triangle\hat{X}(t))^{2}\rangle
  \langle (\triangle\hat{Y}(t))^{2}\rangle \geq \frac{1}{4}$,
where, e.g. $\langle (\triangle\hat{X}(t))^{2}\rangle
 = \langle (\hat{X}(t))^{2}\rangle
-\langle \hat{X}(t)\rangle^{2}$.
Therefore, we can say that the system is able to yield two-mode   squeezing
 if the squeezing factor $S(t)=2\langle (\triangle \hat{X}(t))^{2}\rangle-1<0$  or
$Q(t)=2\langle (\triangle \hat{Y}(t))^{2}\rangle-1<0$. Similar
quantities  can be defined for the single-mode case.

We proceed to the two-mode normal  generating function which is  defined as
\begin{equation}
C^{(W)}_{\cal N}(\lambda,t)= \frac{1}{(\pi\lambda)^{2}}
\int \int\exp(-\frac{1}{\lambda}\sum\limits_{j=1}^{2}|\zeta_{j}|^{2})
C_{\cal N}(\zeta_{1},\zeta_{2},t) d^{2}\zeta_{1}d^{2}\zeta_{2}.  \label{14}
\end{equation}
This function may be used in studying the sum photon-number
distribution and the reduced factorial moments for compound modes.
For the general
term (\ref{12}),  relation (\ref{14}) can be calculated to obtain
\begin{eqnarray}
\begin{array}{lr}
I^{(W)}_{\cal N}(\lambda,t)= \frac{
1}{(1+\lambda\lambda_{+})(1+\lambda\lambda_{-})}
\exp \left[
\frac{A_{+}\lambda}{1+\lambda
\lambda_{+}}
+\frac{A_{-}\lambda}{1+\lambda
\lambda_{-}}\right]\\
\\
\times\exp[i\phi_{1}+i\phi'_{2}
-\frac{1}{2}\sum\limits_{j=1}^{2}(|\alpha_{j}|^{2}
+|\alpha'_{j}|^{2}-2\alpha^{*}_{j}\alpha'_{j})],  \label{15}
\end{array}
\end{eqnarray}
where
\begin{eqnarray}
\begin{array}{lr}
\lambda_{\pm}=\frac{1}{2}[B_{1{\cal N}}(t) +B_{2{\cal N}}(t)]
\pm
\frac{1}{2}
\sqrt{[B_{1{\cal N}}(t) -B_{2{\cal N}}(t)]^{2}+4|D(t)|^{2}},\\
\\
A_{\pm}=\frac{\pm 1}{\lambda_{-}-\lambda_{+}}\left\{
\bar{\alpha}'_{1}(t)\bar{\alpha}'_{2}(t)D(t)
+\bar{\alpha}_{1}(t)\bar{\alpha}_{2}(t)D^{*}(t)\right.\\
\\
-\left.
\bar{\alpha}_{1}(t)\bar{\alpha}'_{1}(t)[B_{2{\cal N}}(t)-\lambda_{\pm}]
-\bar{\alpha}_{2}(t)\bar{\alpha}'_{2}(t)[B_{1{\cal N}}(t)-\lambda_{\pm}]\right\}.
\label{16}
\end{array}
\end{eqnarray}
Expression (\ref{15}) shows that each term of the  generating function
of the system is the two-fold
generating function for Laguerre polynomials.
Further, in each term
the quantities $A_{\pm}$ and $\lambda_{\pm}$ play the role of the
mean numbers of
coherent photons and mean numbers of chaotic photons, respectively.
The sum photon-number
distribution $P(n,t)$  and the reduced factorial moments
$\langle W^{k}(t)\rangle$
 for the compound-mode case
can be defined by means of  the derivative of
$C^{(W)}_{\cal N}(\lambda,t)$ via the relations
\begin{equation}
 P(n,t)=\frac{(-1)^{n}}{n!}\frac{d^{n}}{d\lambda^{n}}
C^{(W)}_{\cal N}(\lambda,t)\Big|_{\lambda=1}, \quad \langle
W^{k}(t)\rangle=(-1)^{k}\frac{d^{k}}{d\lambda^{k}} C^{(W)}_{\cal
N}(\lambda,t)\Big|_{\lambda=0}.\label{17}
\end{equation}
On using (\ref{15}), these quantities can be deduced
to become
\begin{eqnarray}
\begin{array}{lr}
P_{\it I}(n,t)
=
\frac{1
}{(1+\lambda_{-})(1+\lambda_{+})}
\exp\left[ \frac{A_{-}}{1+\lambda_{-}}
+\frac{A_{+}}{1+\lambda_{+}}\right]\\
\\
\times\exp[
i\phi_{1}+i\phi'_{2}
-\frac{1}{2}\sum\limits_{j=1}^{2}(|\alpha_{j}|^{2}
+|\alpha'_{j}|^{2}-2\alpha^{*}_{j}\alpha'_{j})]\\
\\
\times\sum\limits_{l=0}^{n}
\frac{1}{(n-l)!l!}
\left(\frac{\lambda_{-}}{1+\lambda_{-}}\right)^{n-l}
\left(\frac{\lambda_{+}}{1+\lambda_{+}}\right)^{l}
{\rm L}_{n-l}[\frac{A_{-}}{\lambda_{-}(1+\lambda_{-})}]
{\rm L}_{l}[\frac{A_{+}}{\lambda_{+}(1+\lambda_{+})}],
\label{18}
\end{array}
\end{eqnarray}
\begin{eqnarray}
\begin{array}{lr}
\langle
W_{\it I}^{k}(t)
\rangle
=
\exp[
i\phi_{1}+i\phi'_{2}
-\frac{1}{2}\sum\limits_{j=1}^{2}(|\alpha_{j}|^{2}
+|\alpha'_{j}|^{2}-2\alpha^{*}_{j}\alpha'_{j})]\\
\\
\times\sum\limits_{l=0}^{k}
{k \choose l}
\lambda^{k-l}_{-}
\lambda^{l}_{+}
{\rm L}_{k-l}(\frac{A_{-}}{\lambda_{-}})
{\rm L}_{l}(\frac{A_{+}}{\lambda_{+}}), \label{19}
\end{array}
\end{eqnarray}
where
the subscript $I$
 in (\ref{18}) and (\ref{19}) means that these quantities
  have been calculated for the general term
(\ref{12}).  Also in (\ref{18}) and (\ref{19})
$L_{n}(.)$ are the Laguerre polynomials of order $n$ having the form
\begin{equation}
L_{n}(x)=\sum\limits_{m=0}^{n}\frac{(n!)^{2}
(-x)^{m}}{(n-m)! (m!)^{2} }.\label{19a}
\end{equation}
We should mention that the phase terms, e.g.
the term in the
second line of (\ref{18}), exist  only  in the
off-diagonal elements indicating that  these elements are
suppressed  faster than the diagonal elements.  In other words, the
greater the ``distance" between the components of the cat states, the more
rapidly the off-diagonal elements are damped.

On the other hand, the corresponding quantities for the single-mode case
can be obtained with the help of the single-mode normally ordered characteristic
 function and applying  the same procedures as before.
To be more specific,   a general term $I'(\zeta_{1},t)$ of the
first-mode case  can be obtained from
(\ref{12}) by  setting $\zeta_{2}=0$ and consequently the
corresponding photon-number distribution and the reduced factorial moments
 are
\begin{eqnarray}
\begin{array}{lr}
P_{\it I'}(n_{1},t)
=
\exp\left[
i\phi_{1}+i\phi'_{2}
- \frac{ \bar{\alpha}_{1}(t)\bar{\alpha}'_{1}(t)}{
1+B_{1{\cal N}}(t)}
-\frac{1}{2}\sum\limits_{j=1}^{2}(|\alpha_{j}|^{2}
+|\alpha'_{j}|^{2}-2\alpha^{*}_{j}\alpha'_{j})
\right]\\
\\
\times
\left[\frac{B_{1{\cal N}}(t)}{1+B_{1{\cal N}}(t)}\right]^{n_{1}}
\frac{1}{1+B_{1{\cal N}}(t)}
 {\rm L}_{n_{1}}[\frac{
-\bar{\alpha}_{1}(t)\bar{\alpha}'_{1}(t)}{B_{1{\cal N}}(t)(1+B_{1{\cal N}}(t))}],
\label{19b}
\end{array}
\end{eqnarray}
\begin{eqnarray}
\begin{array}{lr}
\langle W_{\it I'}^{k}(t)\rangle = \exp\left[i\phi_{1}+i\phi'_{2}
-\frac{1}{2}\sum\limits_{j=1}^{2}(|\alpha_{j}|^{2}
+|\alpha'_{j}|^{2}-2\alpha^{*}_{j}\alpha'_{j})\right]\\
\\
\times B^{k}_{1{\cal N}}(t)
{\rm L}_{k}[\frac{
-\bar{\alpha}_{1}(t)\bar{\alpha}'_{1}(t)}{B_{1{\cal N}}(t)}].
\label{20}
\end{array}
\end{eqnarray}

Finally, the quantum properties of  the system can  be visualized well
 by  analyzing its Wigner function ($W$).
Actually, this function is sensitive to the interference in phase space
and therefore it is helpful to be examined.
The single-mode   Wigner function
is defined as
\begin{equation}
W(z,t )=\pi^{-2} \int d^{2}\zeta_{1}
\exp(-\frac{1}{2}|\zeta_{1}|^{2})
C_{\cal N}(\zeta_{1},t)  \exp (z\zeta_{1}^{*}-\zeta_{1} z^{*}),
\label{21}
\end{equation}
where $C_{\cal N}(\zeta_{1},t)$ is the single-mode
normally ordered characteristic function.
 The Wigner function for the single-mode general term is
\begin{eqnarray}
\begin{array}{lr}
W_{\it I'}(z,t)=\frac{2}{\pi [1+2B_{1{\cal N}}(t)]}
\exp[
i\phi_{1}+i\phi'_{2}
-\frac{1}{2}\sum\limits_{j=1}^{2}(|\alpha_{j}|^{2}
+|\alpha'_{j}|^{2}-2\alpha^{*}_{j}\alpha'_{j})]\\
\\
\times\exp\left\{-2\frac{
[\bar{\alpha}^{*}_{1}(t)-z^{*}]
[\bar{\alpha}'_{1}(t)-z]}{1+2B_{1{\cal N}}(t)}\right\}.
\label{22}
\end{array}
\end{eqnarray}
 In this expression
$B_{1{\cal N}}(t)$ is always positive and  represents the sum of
 quantum noises in the field, related to quantum
fluctuations in the
interaction and the mean value of  reservoirs oscillators (the number of
photons or phonons
transferred from the reservoirs to the quantum system). This shows that
the width of this part of $W$ function becomes  broader when the coupling
with environment is considered.

In the following
sections we analyze the behaviour of the system under discussion
 on the basis
of the results of the present section. We investigate the properties of both
  the single-mode and  compound-mode  cases.

\section{Results for the single-mode case}
In our analysis for the dissipative case  we consider symmetrical
losses that is
$\gamma_{j}=\gamma$ and $\langle \hat{n}_{jd}\rangle
=\bar{n}$ for $j=1,2$.
 We analyze  two cases:
underdamped  case when $2g>\gamma$  and overdamped  case when
$2g<\gamma$. The origin of these
conditions can  easily  be recognized by careful examination of
the time dependent coefficients in appendix A.
For example, it holds that
\begin{equation}
f_{1}(t)=\frac{1}{2}\left\{\exp [(g-\frac{\gamma}{2})t] +
\exp [-(g+\frac{\gamma}{2})t]\right\}. \label{23}
\end{equation}

In the following we discuss the behaviour of $W$ function, photon-number
distribution, squeezing  and reduced factorial moments
 for the single-mode case.
\subsection{Wigner function}
Investigation of the single-mode $W$ function is important since
it is informative and  sensitive to the interference
in phase space, and also it  gives a prediction to  the possible
occurrence of the  nonclassical effects in the system.
Furthermore, this function can  be obtained experimentally
 using the optical homodyne tomography \cite{tom}.

In general, the $W$ functions of ECS,
 OCS  and YSS (at $t=0$) are consisting of two Gaussian bells,
resulting from statistical mixture of individual composite states and
interference fringes in between (signature of the nonclassical effects)
 originating from the superposition
between different components of the states. Nevertheless, the
locations of the extreme values of the
interference fringes of ECS, OCS and YCS are quite different.
There are several papers, e.g., \cite{buz2,{buz3},{filip},{aga}}
that have been devoted to these fringes  making
them less or more pronounced by allowing  the cat states to evolve in
different quantum optical systems. For example, the decay
rate of the interference fringes of the $W$ function of  cat states
of a mode which is coupled to a phase-sensitive reservoir
(squeezed reservoirs) \cite{buz3}, \cite{luk} p. 114,  can be smaller or larger
than that  for  ordinary thermal heat bath reservoir \cite{buz2}.
Furthermore,  in these systems the
 fringes can be washed out completely for a specific choice
of the interaction parameters.
  In the language of the density matrix this means
that the contribution of the off-diagonal elements is small
compared with the contribution of  the diagonal elements but it is not
necessarily zero. Of course, such diagonalization produces
a statistical mixture of
coherent states, which are  close to classical wavepackets.
Now if we consider  the $W$ function of the signal mode
 for the system under discussion, the situation becomes more complicated
 than before \cite{buz2,{buz3},{filip}} owing to the entanglement
 between the signal and idler modes.
The structure of the density matrix (\ref{8})  carries more
information, e.g., when the signal and idler modes are initially
prepared in a distinguishable macroscopic cat state and after
switching on the
interaction, the initial two Gaussian bells of the $W$ function
of the signal mode can be transformed, in principle, into four-fold form
as indicated in (\ref{9}), and the initial interference fringes
of the cat will be
dramatically changed  during the interaction as a result of the
competition between the different components
of $\hat{\rho}_{\rm AI}(0)$ and $\hat{\rho}_{\rm SI}(0)$.
Moreover, during  evolution, the rate of movement of the centres
of  peaks of the $W$ function  is rather different
 and this leads to irregularity in  its  behaviour.
So that the evolution of the cat states in the present interaction
yields
different types of multicomponent cat states \cite{mog}.

%%%%%%%%%%%%%%%%%%%%%%%%%%%%%%%%%%%%%%%%%%%%%%%%%%%%%%%%%%%%%%%
%%%%%%%%%%%%%%%%%%%%%%%%%%%%%%%%%%%%%%%%%%%%%%%%%%%%%%%%%%%%%%%%%%%
%%%%%%%%%%%%%%%%%%%%%%%%%%%%%%%%%%%%%%%%%%%%%%%%%%%%%%%%%%%%%%%%%%%%%
\begin{figure}[h]%
 \centering
 \subfigure[]{\includegraphics[width=5cm]{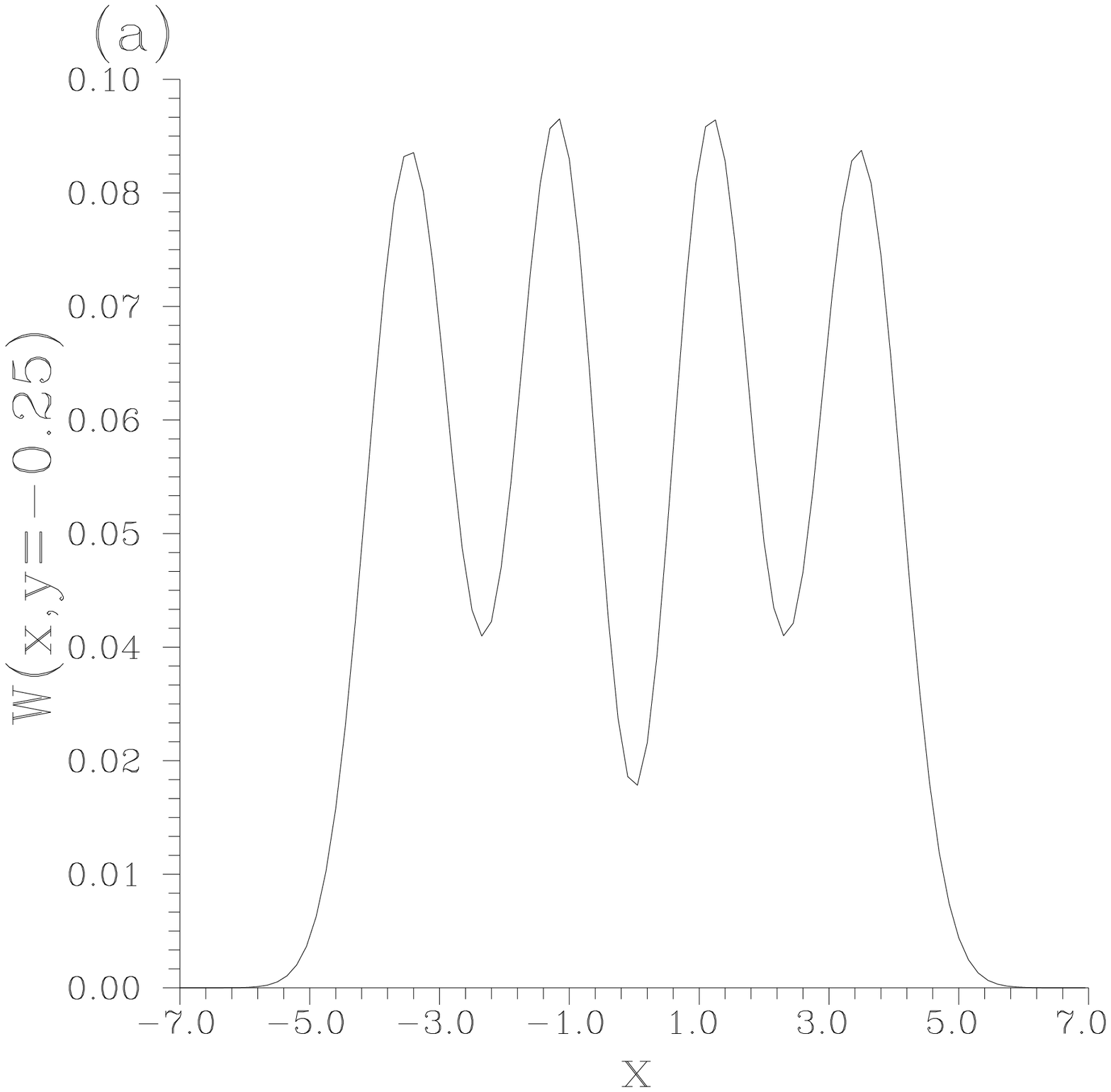}}
\subfigure[]{\includegraphics[width=5cm]{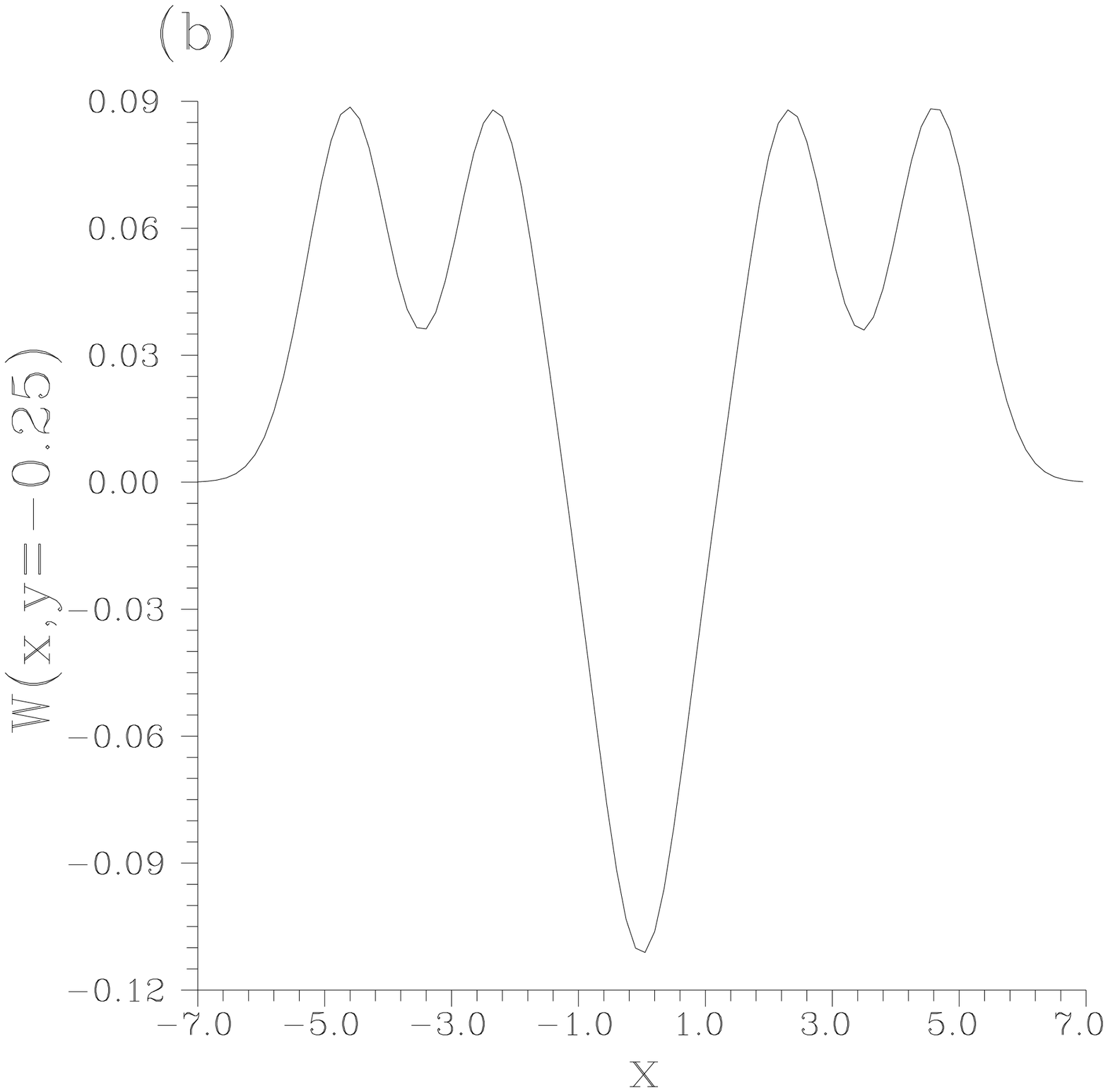}}
\subfigure[]{\includegraphics[width=5cm]{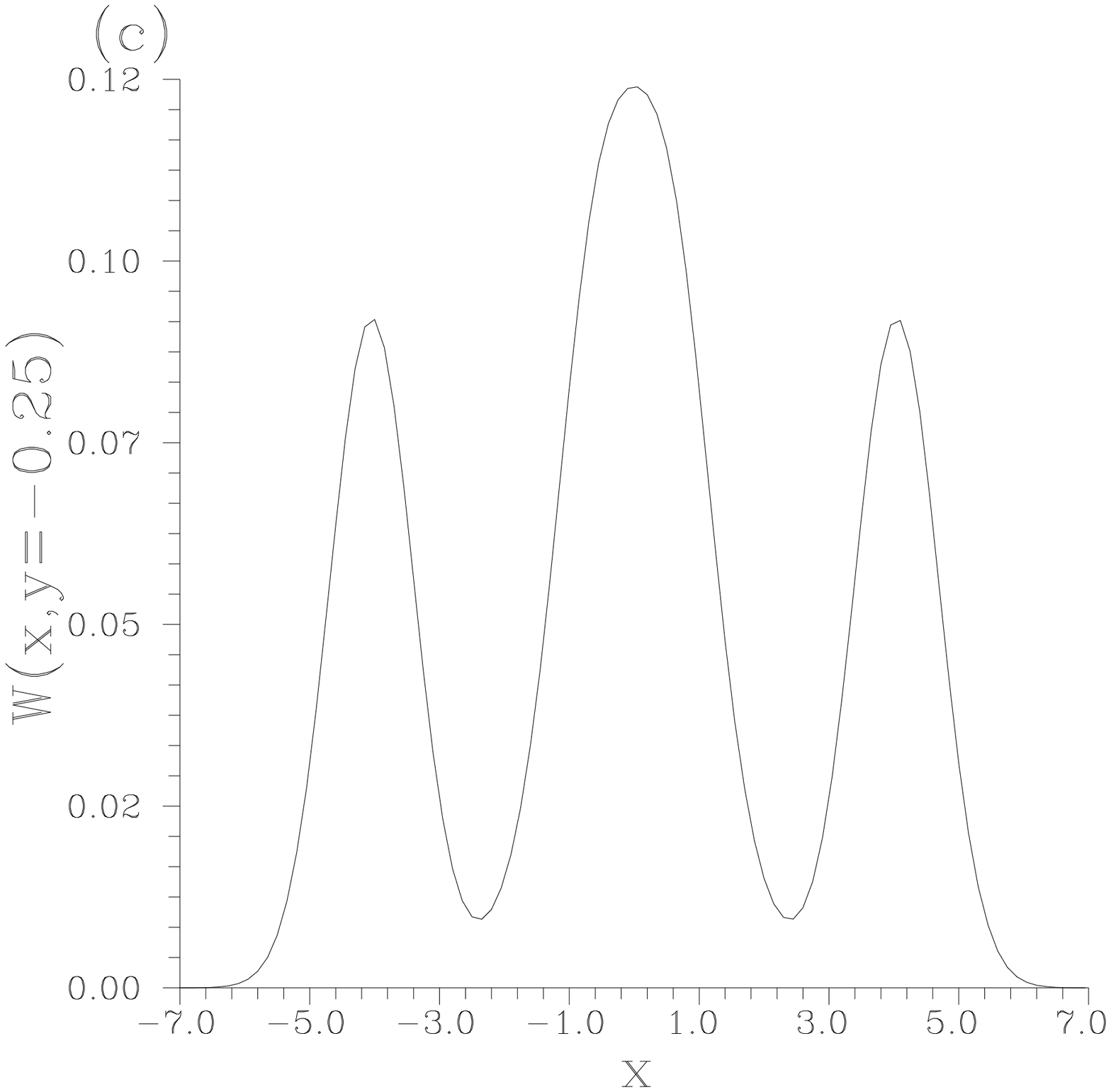}}
 \caption{
The $W$ function of the signal mode when the two modes (signal and
idler) are initially in ECS
 for $t=0.55,g=1,
\psi_{1}=\psi_{2}=0, \gamma=\bar{n}=0$ and $\phi=\frac{\pi}{2}$
and for: a) $(|\alpha_{1}|,|\alpha_{2}|)=(2,2)$; b)
$(|\alpha_{1}|,|\alpha_{2}|)=(3,2)$; c)
$(|\alpha_{1}|,|\alpha_{2}|)=(2,3)$.}
\end{figure}
%%%%%%%%%%%%%%%%%%%%%%%%%%%%%%%%%%%%%%%%%%%%%%%%%%%%%%%%%%%%

All these facts can be seen in Figs. 1a--c, where
 we have displayed the $W$ function
for the signal mode (excluding losses) for the given values of the parameters.
In these figures (and throughout this paper) we have taken
 $\alpha_{j}=|\alpha_{j}|\exp(i\psi_{j})$,
where $\psi_{j}$ is the phase of the initial amplitude of the $j$th mode
and also we have defined $z$ in
(\ref{21}) as $z=x+iy$ where the quadrature variances
$\langle (\triangle \hat{X}(t))^{2}\rangle $ and
$\langle (\triangle \hat{Y}(t))^{2}\rangle$ of the signal mode
are associated with the real part
$x$ and imaginary part $y$, respectively.
It is worth mentioning that  the  $W$ function curves
contain most of their information within one plane, in particular, for
constant values of $y$. For this reason and for simplicity we have
plotted a $W$ function in one
plane considering $y\simeq -0.25$, where the $W$ function
illustrates more pronounced nonclassical effects (e.g., by means
of negative values)
for the case $|\alpha_{1}|>|\alpha_{2}|$ and also it includes a
representative information for the other cases.
Fig. 1a shows  that the initial nonclassical negative values of the
$W$ function have been smeared out and the system collapses to
four-component statistical mixture state.
Further, such behaviour reveals  that  decoherence can be established
via the entanglement between different modes in the parametric
processes.
In these cases the single-mode behaviour undergoes amplification
resulting from  the spontaneous pump photon decay \cite{band}.
It should be borne in mind that  in the present case the system
is completely isolated
and then such type of decoherence can be called nondissipative
 decoherence \cite{lfo}.
On the other hand, the nonclassical negative values can be recovered
by controlling  the ``distance" between the
 initial cat states of the signal and idler modes.
 To be more specific, when $|\alpha_{1}|>|\alpha_{2}|$ these negative
 values can be realized, however, when $|\alpha_{2}|>|\alpha_{1}|$
they will  disappear as indicated in Figs. 1b and c,
respectively.
In Fig. 1c  a three-peak structure is dominant.
Comparison of Figs. 1a,b and c is instructive.
So, excluding the influence of the environment the system can
decohere and recohere by adjusting the initial distance between
the components of the cat  of the input modes.
However, for the dissipative case the decoherence process is related
not only to the amplification of the pumping field but also to
damping of radiation caused by the flux of coherent energy from radiation
to the reservoirs and noise from reservoirs to the radiation.
Under these circumstances the decoherence can be achieved in a
time  shorter than  that for the undamped case.
\setcounter{figure}{1}
%%%%%%%%%%%%%%%%%%%%%%%%%%%%%%%%%%%%%%%%%%%%%%%%%%%%%%%%%%%%%%%
\begin{figure}[h]%
 \centering
  \subfigure[]{\includegraphics[width=8cm]{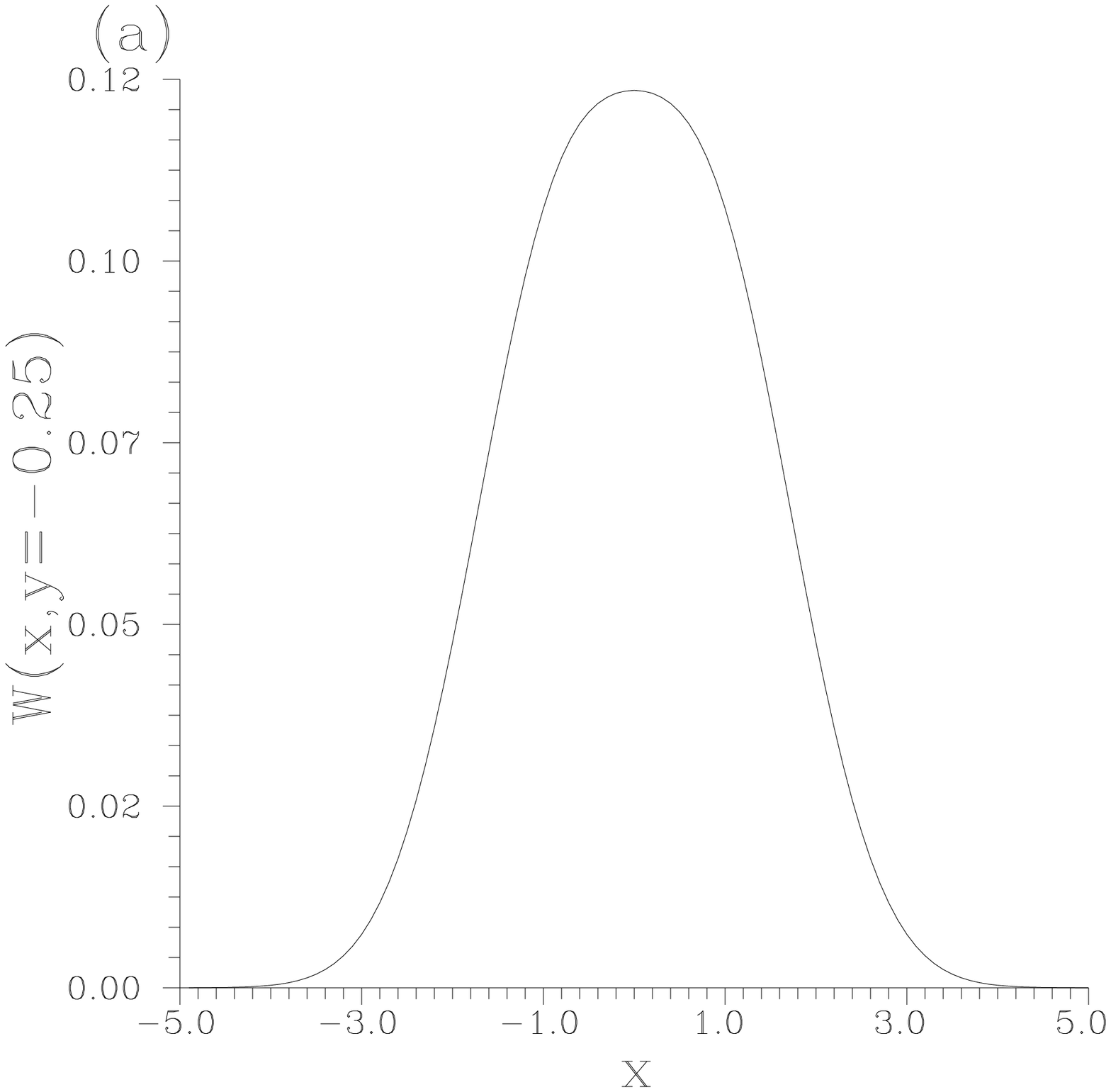}}
\subfigure[]{\includegraphics[width=8cm]{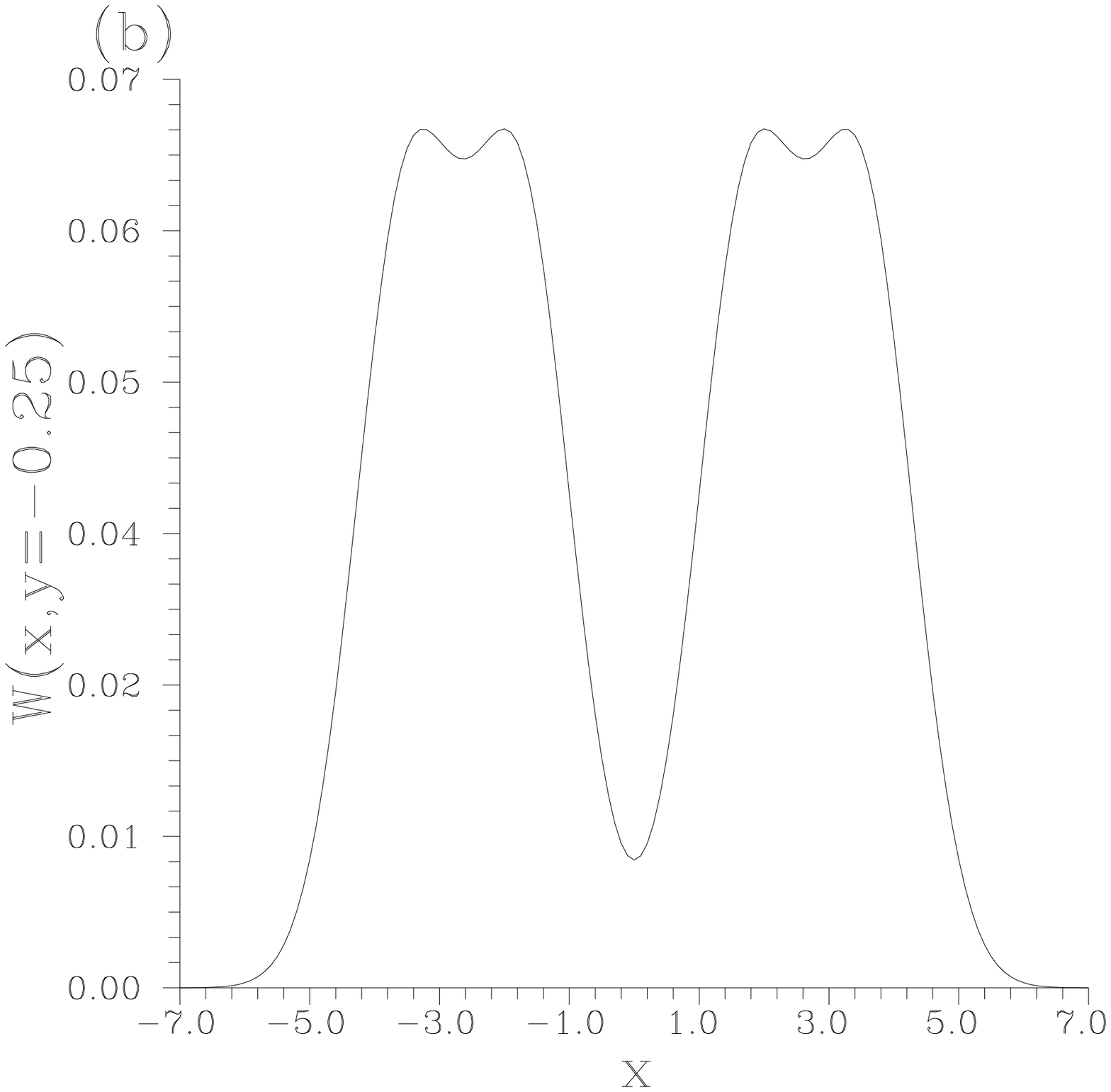}} \caption{ The
$W$ function of the signal mode for the same situation as in Fig.
1b except
 $\bar{n}=1$ and
 a) overdamped  case $\gamma=2g+3$; b) underdamped case $\gamma=2g-1$.}
\end{figure}
%%%%%%%%%%%%%%%%%%%%%%%%%%%%%%%%%%%%%%%%%%%%%%%%%%%%%%%%%%%%
We displayed  the $W$ function for the overdamped  and underdamped cases in
Figs. 2a and b, respectively, for the same situation as in Fig. 1b.
These figures show overall distortion due to the dissipative nature of
dynamics.
It is clear that the origin of the main contribution is in the
diagonal elements of the density matrix and then the
negative values of the $W$ function of the undamped case
 (Fig. 1b) are washed out.
 From Fig. 2a one can observe that the
$W$ function  exhibits the well-known shape for the thermal light,
i.e. the system exhibits Bose-Einstein statistics (super-classical light).
However, from  Fig. 2b (underdamped case) we can observe
that the two-peak structure is  dominant.
This situation is similar to that of the
single harmonic oscillator interacting with the  thermal bath in which
a double Gaussian structure with  missing
oscillatory behaviour occurs \cite{buz2,{buz3},{aga}}.
 The explanation of the behaviour of $W$ function in Fig. 2
can be understood by analyzing the behaviour of the $W$ function
for the single-mode general term (\ref{22}). Let us restrict ourselves
to the overdamped case for which the values of the parameters $\bar{\alpha}_{j}
(t)$ are exponentially decaying (this can be easily checked)  when
the interaction is going on and accordingly the centres of the peaks move
toward the origin and eventually  the peaks merge with  each other.
The opposite situation occurs for the underdamped case.
It is worth mentioning that in these  cases (overdamped and underdamped cases)
 the contribution
of the off-diagonal elements corresponding
to $\hat{\rho}_{\rm AI}(0)$ is suppressed faster than that of
$\hat{\rho}_{\rm SI}(0)$.
This point will be discussed  in the sum photon-number
distribution in the following section.

In general, if the  interaction  between different components in the
system occurs (regardless whether the interaction with  environment is
considered or not),
the  decoherence gradually increases and thus the
 system evolves into a mixture
(${\rm Tr}\hat{\rho}^{2}_{1}(t)<1,\quad \hat{\rho}_{1}(t)$ is the reduced
density matrix of the signal mode), i.e. there  occurs a
destruction of the inherent nonclassical effects.
The rate of   destruction
 is  sensitive to the nature of both the
reservoir  and   parametric processes.
Furthermore, we should mention that the
$P$ function possesses somewhat similar behaviour as
the $W$ function in
such a case, i.e. it can take on negative values in certain
regions for certain values of the parameters and therefore
it cannot behave like a classical
probability distribution function.
Such an effect is independent of the type of cat states,
which are initially used in the
interaction.
Finally, we should stress that it is difficult to analyze the
behaviour of the
off-diagonal elements  to obtain concrete information, so we
have basically concentrated on the computer simulation.

\subsection{Photon-number distribution}
The  concept of photon is an integral part of the modern description of light
and the discrete nature of light
can be demonstrated by a photon detector based on the photoelectric effect.

On the other hand,
one of most interesting nonclassical effects emerging from
the superposition principle is the oscillatory behaviour of
the photon-number distribution.  In general, such behaviour is closely
related to the behaviour of the $W$ function, however, this is necessary
but not sufficient condition.
%%%%%%%%%%%%%%%%%%%%%%%%%%%%%%%%%%%%%%%%%%%%%%%%%%%%%%%%%%%
\begin{figure}[h]%
\includegraphics[width=10cm]{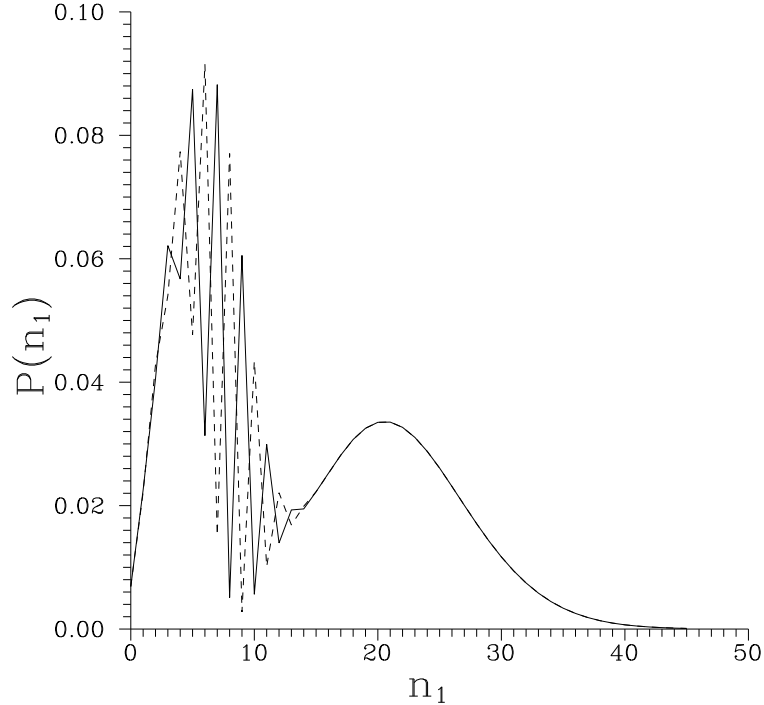} \caption{Single-mode
photon-number distribution of the signal mode when the signal and
idler modes are prepared initially in YSS; the values of the
parameters are the same as in Fig. 1b and
 $\psi =\frac{\pi}{2}$ (solid curve),
$-\frac{\pi}{2}$ (dashed curve).
}
\end{figure}
%%%%%%%%%%%%%%%%%%%%%%%%%%%%%%%%%%%%%%%%%%%%%%%%%%%%%%%%%%%%%%%
For example, the
 photon-number distributions of ECS, OCS and YSS  are completely
different; whereas those  of ECS and OCS exhibit
pairwise oscillations in phase space (even number of
photons can be observed for ECS
and odd numbers for OCS),  the distribution of YSS
is a Poissonian  even though  the behaviour of the  $W$ function
for these states is
qualitatively similar.
In the  interaction under discussion for undamped
case the oscillatory behaviour in the photon-number distribution can be
established  even if the initial cat states exhibit Poissonian
statistics. This is of course based on the values of the interaction parameters.
The origin of such behaviour is in the interference in phase space where
 the photon-number distribution of input coherent light is always
displaying a single-peak structure, which  is broader than the corresponding
Poisson distribution with the same mean photon number.

We  start our discussion by investigating the behaviour of the undamped case.
We consider here the photon-number distribution
$P(n_{1})$ of the signal mode
when the signal and idler modes are  initially prepared
in the YSS.
We have seen that this quantity
can exhibit oscillatory behaviour after switching on of the
interaction by a suitable time provided that $|\alpha_{1}|> |\alpha_{2}|$,
as indicated in Fig. 3 for shown values of
the parameters. In this figure  $\psi=\phi-\psi_{1}-\psi_{2}$, where
$\psi_{j}$ is  the phase of the initial  $j$th mode (signal or idler) and
$\phi$ is the phase of the pump as before. In this case $\psi$
represents  the phase mismatch.
Furthermore,  one can see that when
$\psi$ changes from $\pi/2$ to $-\pi/2$ the parity of oscillations
changes (compare dashed and solid curves). The reason for taking
 $\psi=\pm\pi/2$ can be found in \cite{faisal}.
The origin of
the oscillatory behaviour in the photon-number distribution $P(n_{1})$ here
is the competition between the contributions of $\hat{\rho}_{\rm M}(0)$
and of $\hat{\rho}_{\rm SI}(0)$, as we will see in the sum photon-number
distribution.
%%%%%%%%%%%%%%%%%%%%%%%%%%%%%%%%%%%%%%%%%%%%%%%%%%%%%%%%%%%%%%%%%%%
\begin{figure}[h]%
 \centering
  \includegraphics[width=10cm]{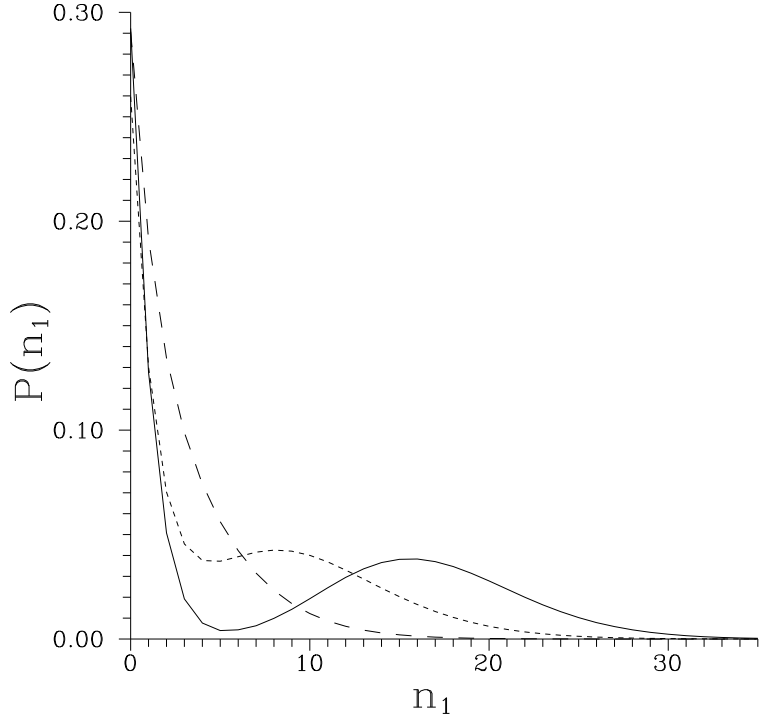}
 \caption{Single-mode photon-number distribution of the signal
mode when the signal and idler modes are prepared initially in
YSS, the values of the parameters are the same as in Fig. 1c with
$\psi =\frac{\pi}{2}$ for $ \gamma=\bar{n}=0$ (undamped case-solid
curve); $\bar{n}=1,\gamma=2g-1$ (underdamped case-short-dashed
curve); $\bar{n}=1,\gamma=2g+1$ (overdamped case-long-dashed
curve). }
\end{figure}
%%%%%%%%%%%%%%%%%%%%%%%%%%%%%%%%%%%%%%%%%%%%%%%%%%%%%%%%%%%%%%%
Further, it is worth mentioning that the behaviour of
$P(n_{1})$ in Fig. 3 is similar to that of initially cat state of a
mode coupled to a phase-sensitive reservoir
\cite{buz3}, \cite{luk} p. 114, however, the source of the oscillations
in \cite{buz3} is the
phase information included in the reservoir, which can be
transferred to the field.
In Fig. 4 we have displayed $P(n_{1})$ for the same situation as given
by the solid curve in
Fig. 3, but for $|\alpha_{1}|<|\alpha_{2}|$
(solid curve); further the damped cases are considered: underdamped case
(short-dashed curve) and overdamped case (long-dashed curve).
Comparing the solid curves in Figs. 3 and 4, we see from Fig. 4 that
 the oscillations in  $P(n_{1})$ are
smoothed out.  Further, the comparison of various curves in Fig. 4
shows that the behaviour of   $P(n_{1})$ for underdamped and undamped cases
is similar in the sense that they include smooth oscillations.
These smooth oscillations are  completely washed out for
the overdamped case, as is expected.
Actually, for $|\alpha_{2}|>|\alpha_{1}|$  coherence is lost and the main
contribution is related to the energy of the field mode (diagonal terms).
As we can see the behaviour of the photon-number distribution is in a good
agreement with  that of $W$ function.
Finally, it has been verified that the behaviour of $P(n_{1})$
for the damped case when
 $|\alpha_{1}|>|\alpha_{2}|$ and $|\alpha_{1}|<|\alpha_{2}|$
 is quite similar.
This is connected with the fact
that the oscillatory behaviour in the photon-number distribution
is highly sensitive to the dissipation dynamics.

\subsection{Single-mode squeezing and reduced factorial moments}
Squeezing is one of the most important phenomena in quantum optics
because  of its  applications in various areas, e.g.
in optics communication, quantum information theory, etc.
\cite{inf1}. Squeezed light can be measured by a homodyne detection in
which the signal is superimposed on a strong coherent beam of the local
oscillator.
Furthermore, quite recently it has been shown experimentally that there
is an evidence of  squeezed light in the biological systems \cite{popp}.
So an analysis of  squeezing phenomenon in quantum optical systems
is an important
 topic.

 We start our investigation by determining the behaviour of the single-mode
squeezing.
Generally, in the system under consideration the output mode loses its
initial squeezing feature
during the interaction \cite{dom} as a result of an amplification process
 and the interaction with the environment, which
   accelerates the loss in quantum fluctuations.
Here we analyze the influence of different types of cat states on the behaviour
of quadrature squeezing.
To do so, considering $\alpha_{j}$  to be real, we
 write down  the quadrature variance
 of the $Y$-component of(which is
expected to yield squeezing) for the signal mode and various
initial input cat states as follows:
\begin{equation}
 Q_{\rm ee}(t)=\frac{1}{2}\left\{ B_{1{\cal N}}(t)
+\alpha_{1}^{2}f_{1}^{2}(t)(\tanh\alpha_{1}^{2} -1)
+\alpha_{2}^{2}|f_{2}(t)|^{2}[\tanh\alpha_{2}^{2} +\cos (2\phi)]
\right\},\;\; \label{24}
\end{equation}
\begin{equation}
 Q_{\rm oe}(t)=\frac{1}{2}\left\{ B_{1{\cal N}}(t)
+\alpha_{1}^{2}f_{1}^{2}(t)(\coth\alpha_{1}^{2} -1)
+\alpha_{2}^{2}|f_{2}(t)|^{2}[\tanh\alpha_{2}^{2} +\cos (2\phi)]
\right\}, \;\; \label{25}
\end{equation}
\begin{eqnarray}
\begin{array}{lr}
Q_{\rm ey}(t)=\frac{1}{2}\left\{
B_{1{\cal N}}(t)
+\alpha_{1}^{2}f_{1}^{2}(t)(\tanh\alpha_{1}^{2} -1)\right.
\\
\\
+\left.\alpha_{2}^{2}|f_{2}(t)|^{2}[1 +\cos (2\phi)
-2\exp(-4\alpha_{2}^{2})\sin^{2}\phi]
\right\}
.\label{26}
\end{array}
\end{eqnarray}
In these expressions  the subscripts {\rm ee}, {\rm oe}
and {\rm ey} stand for the initial (signal, idler) modes which are
in (ECS, ECS), (OCS, ECS) and (ECS, YSS), respectively. Furthermore,
one can note that the significant value of squeezing can be
obtained when the pump phase is $\phi=\pm\pi/2$.
Also it can be mentioned that the correlation between signal and idler modes
does not occur  since  cross terms such as $\alpha_{1}\alpha_{2}$ are absent.
Actually, this is connected with the two-photon nature of ECS and OCS
where $\langle \hat{a}^{m}_{j}(0)\rangle=0, j=1,2$, when $m$ is  an odd integer.
It is worth mentioning that the origin of losses caused by
reservoirs in expressions (\ref{24})--(\ref{26}) is the mean photon
number.
Further, it should be reminded that ECS and YCS
can exhibit  normal squeezing, which is more pronounced for ECS,
however, OCS are unsqueezed states \cite{buz1}.

Further, for the undamped case with $\phi=\pi/2$, expression (\ref {25})
can be rewritten as
\begin{equation}
 Q_{\rm oe}(t)=\frac{1}{2}\left\{ \sinh^{2}(gt)[1
+\alpha_{2}^{2}(\tanh\alpha_{2}^{2} -1)]
+\alpha_{1}^{2}\cosh^{2}(gt)(\coth\alpha_{1}^{2} -1)
\right\}.\label{27}
\end{equation}
The last term in this expression is non-negative and can become zero by
appropriately choosing the value of $\alpha_{1}^{2}$.
In order to  get squeezing in the signal mode,
expression (\ref{27}) must be negative  and  this depends on
the behaviour of the function $f(x)=x(\tanh x -1)$.
In other words,  squeezing can be established
if $1+f(x)<0$, however,
 $-0.3\preceq f(x)<0$,
and consequently
squeezing cannot be obtained.
More specifically,
 to obtain single-mode squeezing from  this device,
the mode under consideration should be prepared
initially in a squeezed-cat state regardless of the type of the
cat state  in
the free port (if it is squeezed or not).
Furthermore, one can easily estimate  how long
the  single-mode squeezing of the initial
light can survive if this light is imposed at the input
of  the parametric amplifier.
Restricting ourselves to the input ECS  and using
 the fact that squeezing is surviving if
$Q_{\rm ee}(t)<0$, the time range  over which such a situation occurs is
\begin{equation}
t<\frac{1}{g}\sinh ^{-1}\sqrt{\frac{-f(\alpha^{2}_{1})}{
1+f(\alpha^{2}_{1})+f(\alpha^{2}_{2})}},
\label{28}
\end{equation}
where $f(x)$ has the same expression as before. For $Q_{\rm oe}(t)$
 (i.e. (\ref{25})) the procedure shows that the required time is a
 complex number and this agrees with the previous remark.

From (\ref{24}) and (\ref{26}) one can easily verify
that the amount of  squeezing available in $Q_{\rm ee}(t)$ is much larger
than that in $Q_{\rm ey}(t)$ provided that $\alpha_{2}$ is finite;
nevertheless,
when the value of $\alpha_{2}$ is zero or large enough, both quantities
are typical.
This implies that the larger the degree of squeezing
in the free port is, the more  squeezing will be available in the mode under
consideration.

%%%%%%%%%%%%%%%%%%%%%%%%%%%%%%%%%%%%%%%%%%%%%%%%%%%%
\begin{figure}[h]%
 \centering
  \includegraphics[width=10cm]{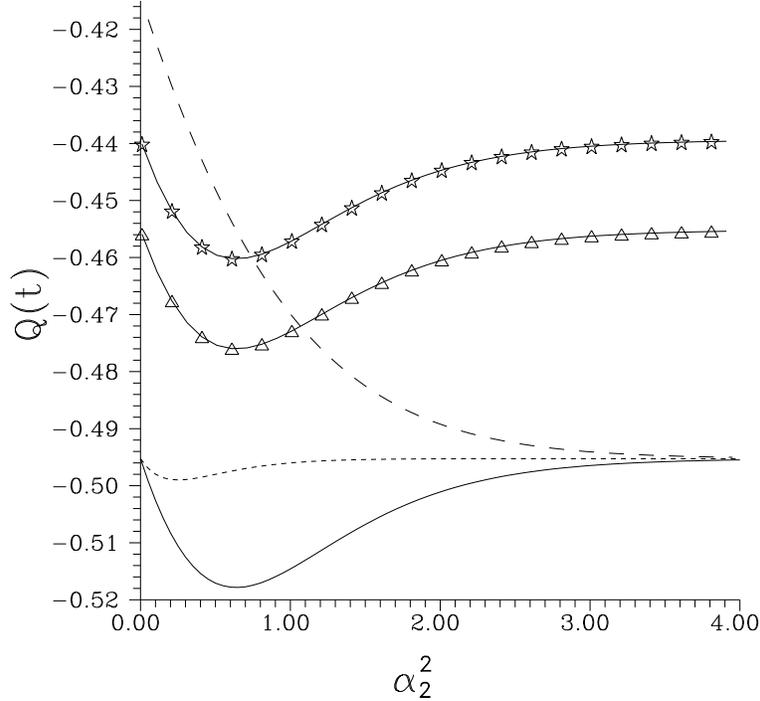}
 \caption{The single-mode squeezing factor $Q(t)$ for the signal
mode
 being initially in ECS and the idler mode  being in ECS (solid curve), YSS
 (short-dashed curve)
and in OCS (long-dashed curve) for $t=0.2, g=1,\alpha^{2}_{1}= 0.7,
\psi_{1}=\psi_{2}=0, \gamma=\bar{n}=0$ and $\phi=\frac{\pi}{2}$.
Triangle- and star-centred curves
are given  for the squeezing factors of
  a zero-temperature heat bath ($\bar{n}=0$) and
a nonzero-temperature heat bath ($\bar{n}=0.1$), respectively,
for the underdamped case with $\gamma=2g-1.6$  and
for the same situation as  represented by the solid curve.
}
\end{figure}
%%%%%%%%%%%%%%%%%%%%%%%%%%%%%%%%%%%%%%%%%%%%%%%%%%%%%%%%
Fig. 5 has been displayed to illustrate the behaviour of the $Q$-squeezing
factor
for the undamped case when
 the signal mode is prepared initially in ECS and the idler mode
in ECS (solid curve), YSS (short-dashed curve) and OCS (long-dashed curve)
in dependence on $\alpha_{2}^{2}$ for the given values of the parameters.
Furthermore, the triangle- and star-centred curves represent the
underdamped case  with a zero-temperature ($\bar{n}= 0$)
 and a nonzero-temperature heat baths ($\bar{n}\neq 0$), respectively,
when the two modes are in ECS.
The choice of  $\alpha^{2}_{1}= 0.7$ is related to the fact
that the initial ECS gives maximum squeezing at this value.
Now the analytical facts discussed above are remarkable in Fig. 5.
Further, from this figure we can also see that
 for the OCS-idler-mode   squeezing is minimum when $\alpha^{2}_{2}$ is close
to zero (indeed there is a singularity at $\alpha_{2}=0$ related to the
nature of the OCS),
increasing monotonically   as $\alpha_{2}$ increases, then it stagnates at
large values of $\alpha_{2}$ yielding its maximum value.
Actually, the behaviour of this case  is quite different from the
behaviour of
the ECS-idler- and YSS-idler-mode cases (compare long-dashed curve
with solid and short-dashed curves in Fig. 5), where
 the maximum value of the former is the minimum value for the latter.
This means
that for large values of $\alpha^{2}_{2}$  the signal mode can produce
the same
value of squeezing regardless of the type of the input cat states in the free
port.
Also this figure shows how one can control the single-mode squeezing
relying on the type of initial cat states in the free port.

We now turn our attention to the damped case.
As we mentioned earlier $B_{1{\cal N}}(t)$ in  expressions
(\ref{24})--(\ref{26})  is always positive and therefore the coupling of
the system with the environment degrades
the amount of the single-mode squeezing.
  We have given two examples in Fig. 5 for the
underdamped case (triangle- and star-centred curves).
Comparing these curves with the solid
one in the same figure, the conclusion becomes clear.
Moreover, from the comparison between triangle-centred curve
and star-centred
curve  we can conclude that in
nonzero-temperature heat bath  case the quantum coherence (i.e.
nonclassical effects) is
lost much faster than in zero-temperature heat bath case
\cite{buz2}.
On the other hand,  the occurrence of squeezing in the damped case
has been verified also in the behaviour of the $W$ function
where for the same values of the parameters we obtained that
it  always exhibits  noise-ellipse  forms of cuts together with
single-peak or two-peak structure according to $\alpha_{2}$
is small or large,
respectively.
It is worth mentioning that
 for the cat states interacting with heat bath the authors of \cite{buz1}
claimed that  squeezing is much more robust
with respect to damping than the oscillations in the
photon-number distribution
or the interference in phase space described by $W$ function.
They obtained this conclusion by analyzing these
 quantities graphically. Nevertheless, in their analysis, squeezing has
been obtained for a specific region of $\alpha$ (in particular when
$\alpha$ is small) whereas the
behaviour of the photon-number distribution and the $W$ function
has been analyzed in a different region (when $\alpha=2$, where  squeezing
does not occur,  see Fig. 8a in \cite{buz1}).
This leads to the incorrect conclusion.
We have examined the behaviour of both the photon-number
distribution and the $W$ function for these cases
using the same values of the parameters as
in \cite{buz1}, however, we have taken $\alpha=1$ (where squeezing
 is noticeable)
and found that the behaviour of these quantities
reflects  equally the properties of squeezed light. For more details
reader can consult \cite{faisall}.

Finally, for the single-mode
reduced factorial moments, which can be measured by a set of photodetectors,
we have found that the behaviour of these  quantities  is
in a good  agreement with the behaviour of the single-mode squeezing.

%%%%%%%%%%%%%%%%%%%%%%%%%%%%%%%%%%%%%%%%%%%%%%%%%%%%%%%%%%%%%%%%%%%%%%
%%%%%%%%%%%%%%%%%%%%%%%%%%%%%%%%%%%%%%%%%%%%%%%%%%%%%%%%%%%%%%%%%%%%%
\section{Results for the compound-mode case}
%%%%%%%%%%%%%%%%%%%%%%%%%%%%%%%%%%%%%%%%%%%%%%%%%%%%%%%%%%%%%%%%%%%%%
%%%%%%%%%%%%%%%%%%%%%%%%%%%%%%%%%%%%%%%%%%%%%%%%%%%%%%%%%%%%%%%%%%%%%
In this section we demonstrate the two-mode properties for the system
under discussion by determining the sum photon-number distribution,
two-mode squeezing and reduced factorial moments.
%%%%%%%%%%%%%%%%%%%%%%%%%%%%%%%%%%%%%%%%%%%%%%%%%%%%%%%%%%%%
\subsection{Sum photon-number distribution}
%%%%%%%%%%%%%%%%%%%%%%%%%%%%%%%%%%%%%%%%%%%%%%%%%%%%%%%%%%%%%
The oscillations in the joint
photon-number distribution might be observed in experiments that
generate two-mode squeezed light in which the two modes may be
distinguished by frequency or by propagation direction. Having separated
the two modes, as was done by using a polarizing beamsplitter \cite{ayt},
 one can send them
directly onto separate photocounters and then build up the joint
photon-number distribution from the photocount statistics of a sequence
of pulses \cite{caves}.

It has been shown \cite{faisal} that the sum
photon-number distribution
$P(n)$ of the nondegenerate parametric amplifier, when the modes
are initially prepared
in coherent state and under certain conditions, displays two regimes,
which exhibit either single-peak structure or
oscillatory behaviour.
%%%%%%%%%%%%%%%%%%%%%%%%%%%%%%%%%%%%%%%%%%%%%%%%%%%%%%%%%%%%%%%
\begin{figure}[h]%
 \centering
 \subfigure[]{\includegraphics[width=8cm]{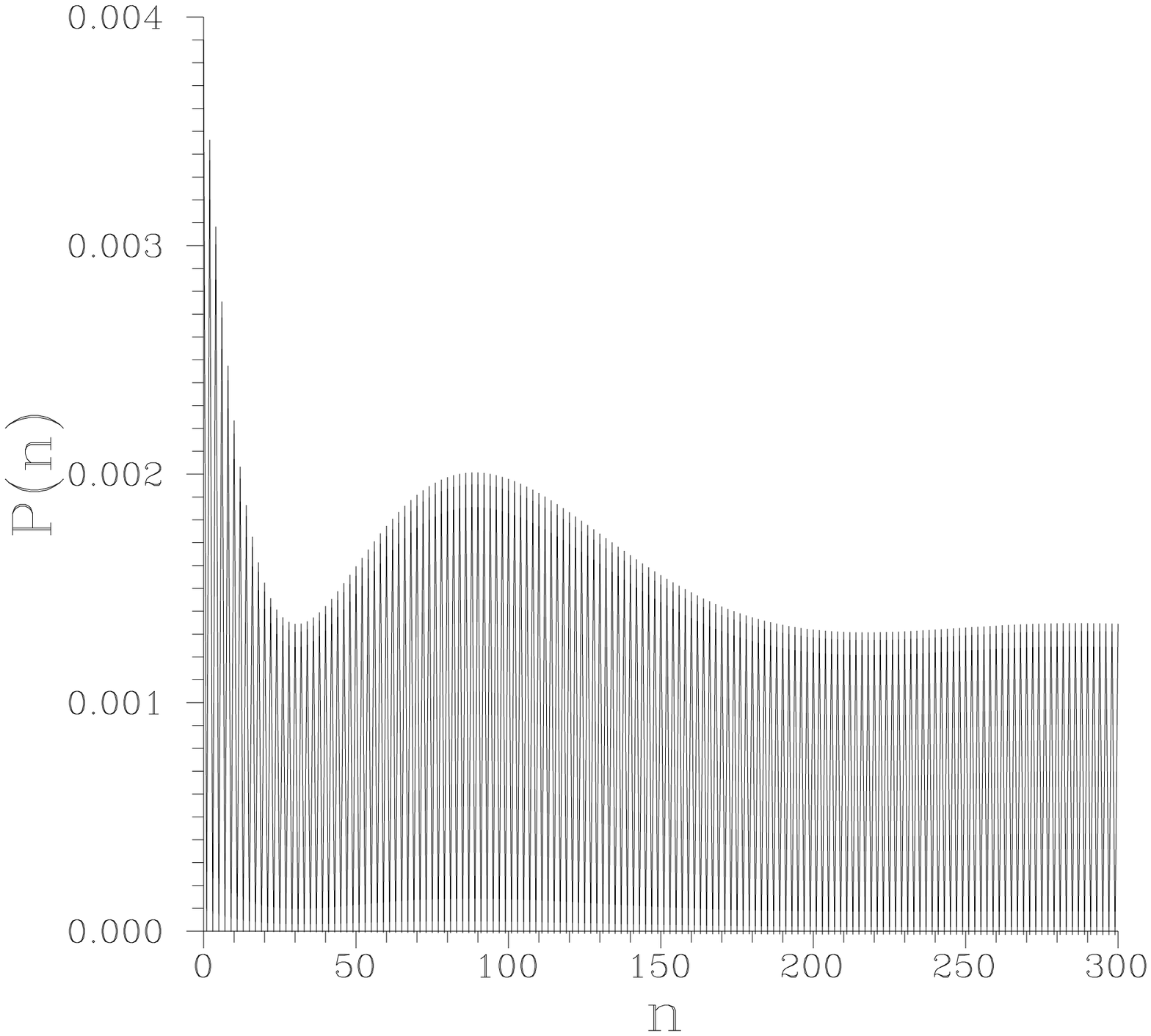}}
    \subfigure[]{\includegraphics[width=4cm]{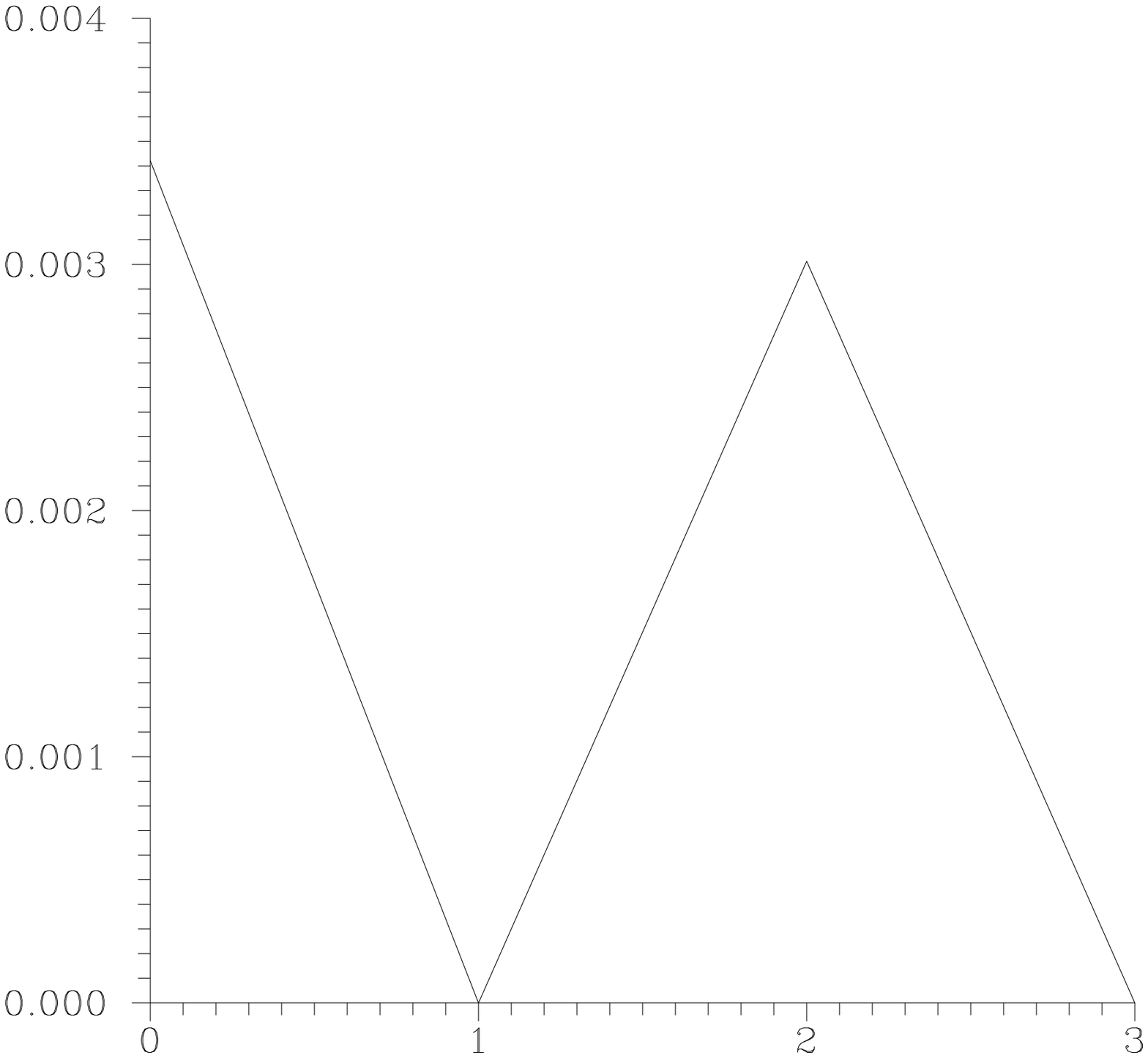}}
  \caption{Sum photon-number
distribution when the signal and idler modes are prepared
initially in ECS with $|\alpha_{1}|=3,|\alpha_{2}|=2, g=10^{4},
t=3\times 10^{-4}, \psi=\frac{\pi}{2}$. On the right corner of the
figure the scheme
 shows the parity of the photon-number sum.
}
\end{figure}
%%%%%%%%%%%%%%%%%%%%%%%%%%%%%%%%%%%%%%%%%%%%%%%%%%%%%%%%%%%%%%%
The single-peak structure can
be narrower (or broader) than that of the corresponding
Poissonian distribution showing nonclassical (or classical) effects.
The striking feature is  that the $P(n)$ can
exhibit, for certain choice of the interaction parameters,
collapses and revivals
in the photon-number domain  somewhat similar
to those known in the JCM.
For the present system the situation is rather complicated  regarding to the
evolution of the density matrix (\ref{8}) where the interference in phase
space is established. Excluding dissipation and using $\psi=\pm\pi/2$
as in \cite{faisal}, we have seen  generally
that $P(n)$ exhibits always oscillatory behaviour irrespective of which
type of cat states
has been considered initially. Moreover, $P(n)$ can yield pairwise
oscillations identifying that  even or odd photons are being observed.
Figs. 6 and 7 have been plotted to show such a phenomenon for given values
of the parameters. From Fig. 6 we observe that the long scale
oscillations in the
behaviour of $P(n)$ with $P(2n+1)=0$ are somewhat similar to those of squeezed
states \cite{yun2}.
Actually for squeezed states  the pairwise oscillations
are explained as  a direct consequence  of the quadratic, or two-photon
nature of the squeeze operator $\hat{S}(r)$, i.e. $
\hat{S}(r)=\exp
[ \frac{r}{2}(\hat{a}^{2}
- \hat{a}^{\dagger2}) ]$  \cite{yun2}.
However, here the origin of the pairwise oscillations
in $P(n)$ is in the competition between the processes described
by three parts of the
density matrix of the field.
%%%%%%%%%%%%%%%%%%%%%%%%%%%%%%%%%%%%%%%%%%%%%%%%%%%%%%%%%%%%%%%
\begin{figure}[h]%
 \centering
 \subfigure[]{ \includegraphics[width=5cm]{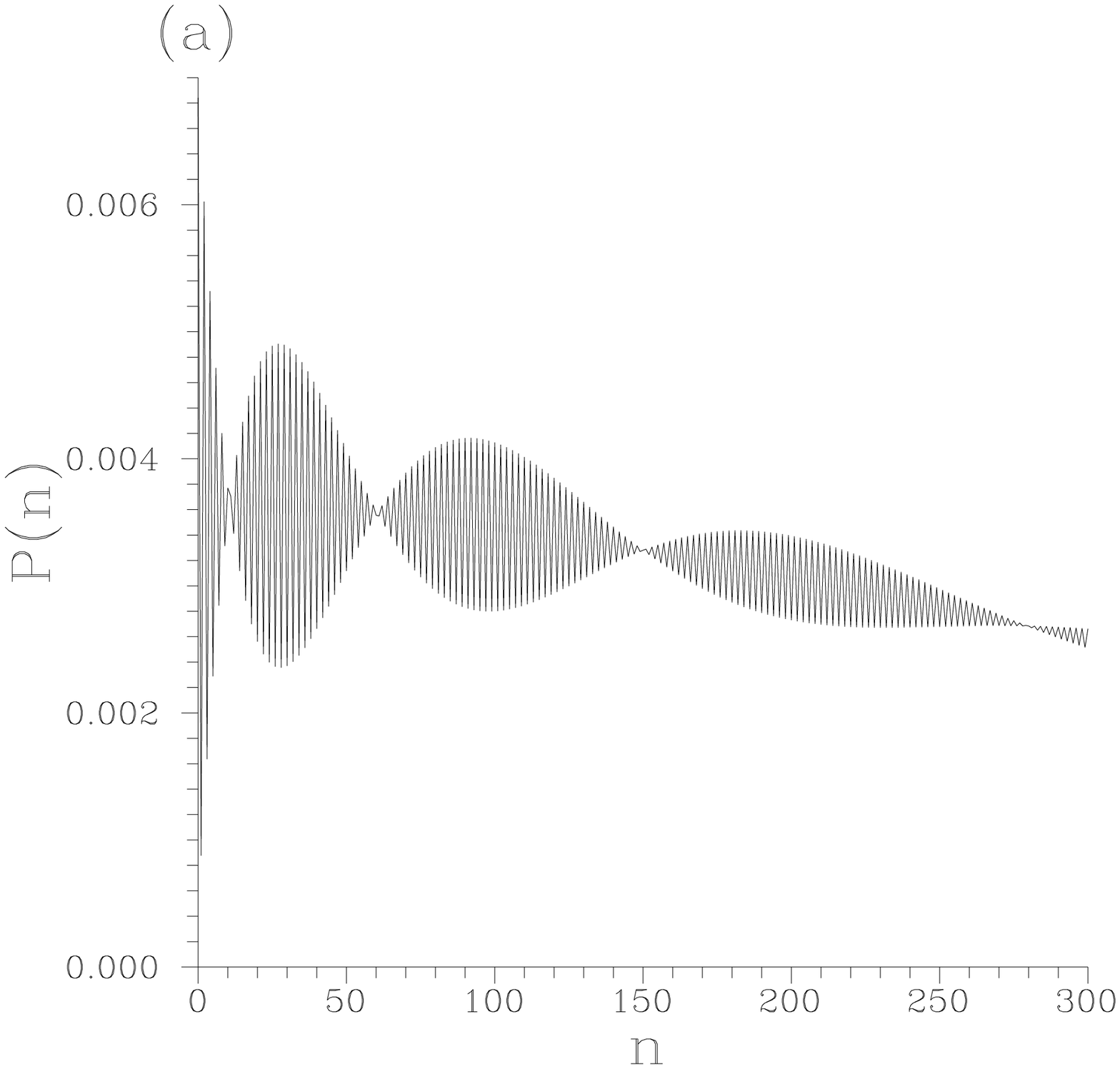}}
\subfigure[]{ \includegraphics[width=5cm]{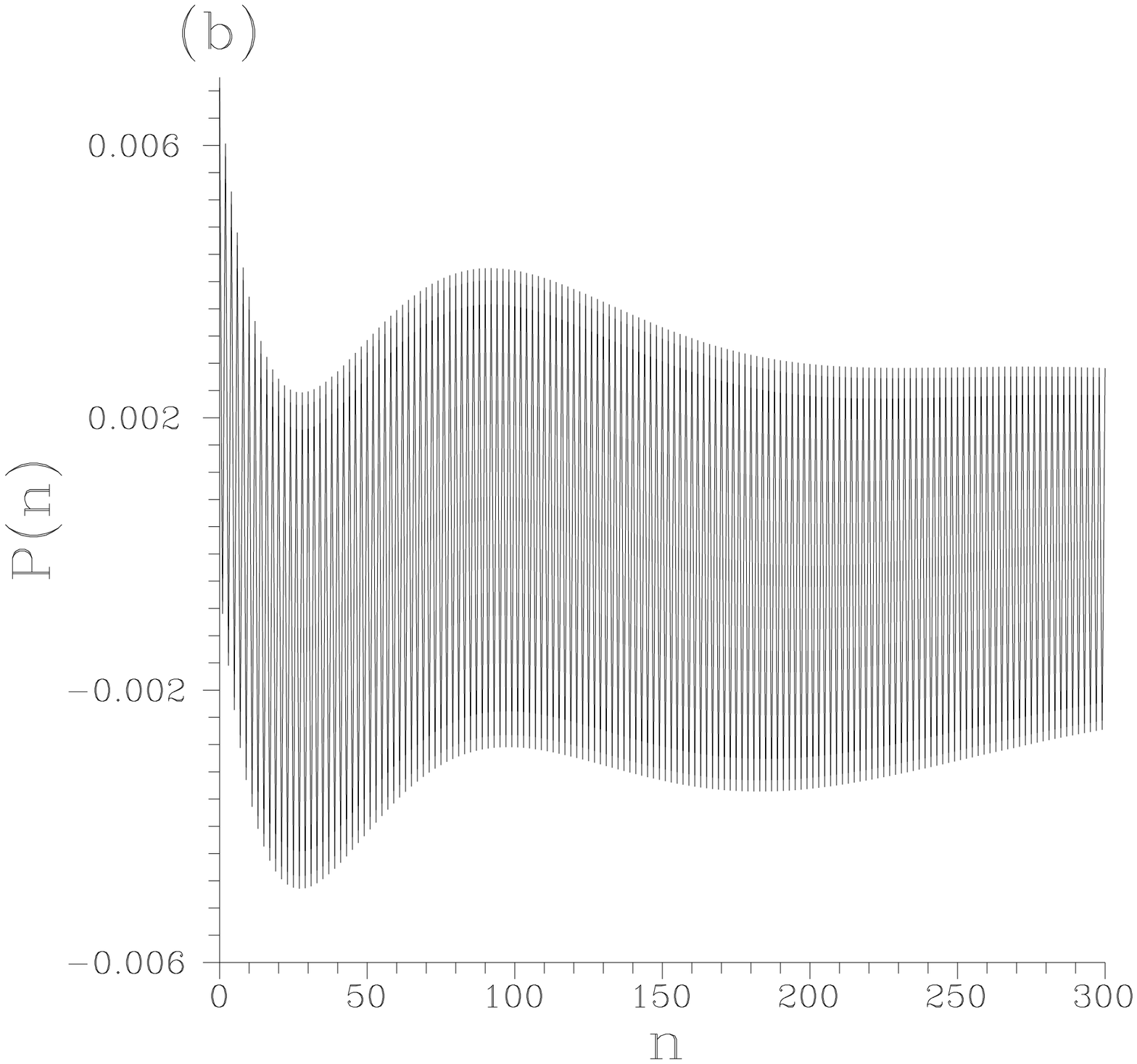}}
\subfigure[]{ \includegraphics[width=5cm]{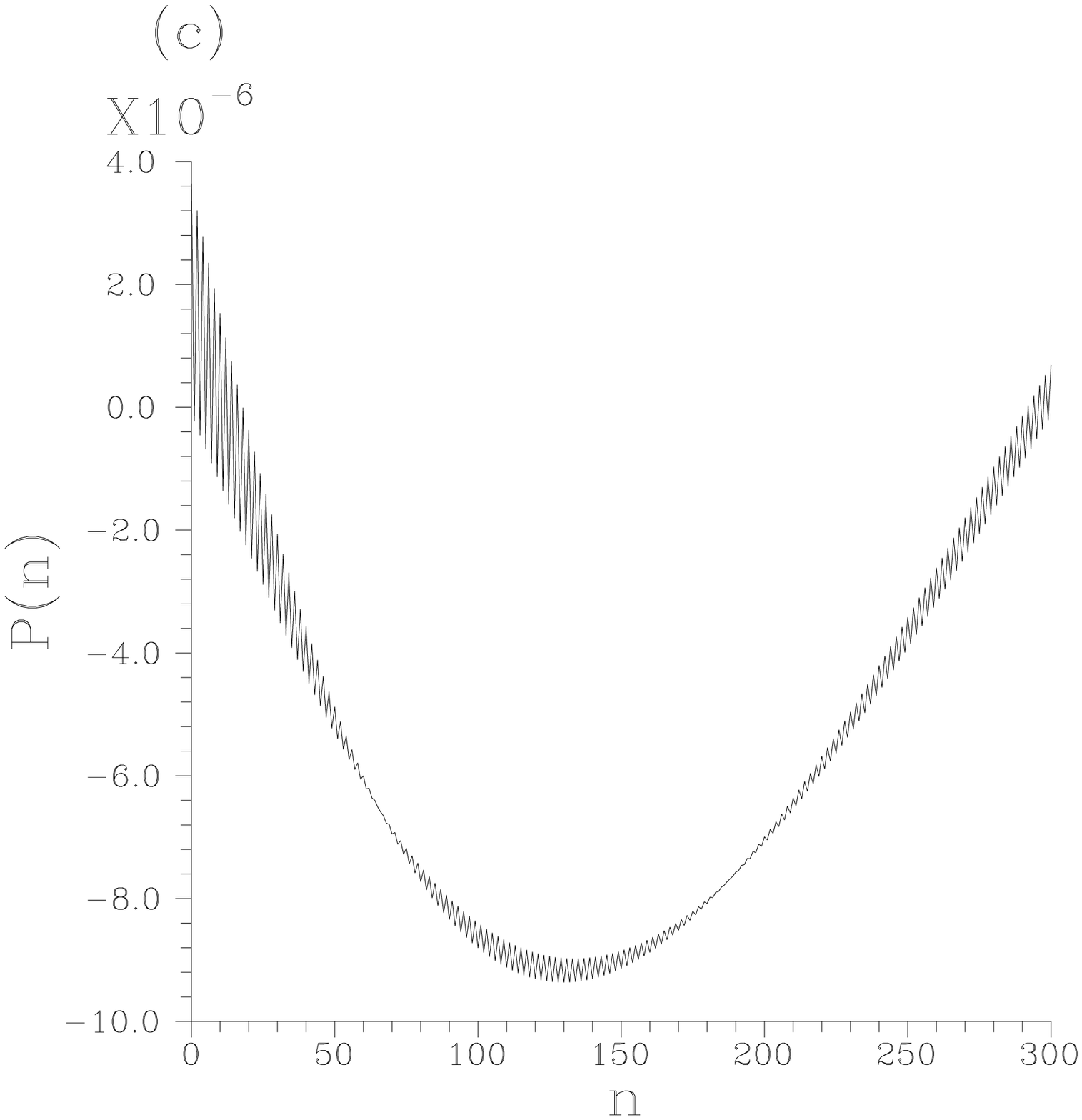}} \caption{Sum
photon-number distributions  corresponding to the three parts of
the density matrix for the same values of the parameters  as in
Fig. 6 and for: a) $P(n)$ resulting from $\hat{\rho}_{\rm M}(0)$;
b) $P(n)$ resulting from $\hat{\rho}_{\rm SI}(0)$; and c) $P(n)$
resulting from $\hat{\rho}_{\rm AI}(0)$. }
\end{figure}
%%%%%%%%%%%%%%%%%%%%%%%%%%%%%%%%%%%%%%%%%%%%%%%%%%%%%%%%%%%%%%%
Figures 7a--c give insight into this point,
i.e. they show the manner in which the photon-number distributions
of the three parts
of the density matrix (\ref{8}) compete.
More precisely, Figs. 7a,b and c represent the sum photon-number
distributions
$P_{\rm M}(n), P_{\rm SI}(n)$ and $P_{\rm AI}(n)$
 associated with $\hat{\rho}_{\rm M}(0), \hat{\rho}_{\rm SI}(0)$ and
 $\hat{\rho}_{\rm AI}(0)$, correspondingly.
 Fig. 7a shows
revival-collapse pattern, which is  resulting from the statistical
mixture part (\ref{9}). Furthermore,  from this figure one can observe
 that when the number of
photons increases, the amplitude of the revivals diminishes, but the revival
periods extend. It is worth mentioning that such a quantum collapse-revival
phenomenon  has been seen for
the photon-number distribution of single-mode
\cite{dutta} and two-mode \cite{mary} squeezed coherent states
with complex squeeze and displacement parameters.
Associations with the quantum phases of the modes may be considered
\cite{mary,sm97}, \cite{luk} p. 281.
From Fig. 7b it is clear that $P_{\rm SI}(n)$
oscillates between negative and positive values.
Comparison of Fig. 7a, Fig. 7b and Fig. 7c shows that
the values of $P_{\rm AI}(n)$ are approximately negligible
compared with those of $P_{\rm M}(n)$ and $P_{\rm SI}(n)$.
Now if we turn our attention back to Fig. 6 we can recognize that
 the competition between
$P_{\rm M}(n)$ and $P_{\rm SI}(n)$ leads
 to the destruction of collapse-revival phenomenon (which appeared
 in Fig. 7a), however,
the distribution ``evolves" in an interesting manner similarly as that for
squeezed states.
In fact, the situation here is in contrast with that of the
evolution of cat states
in the JCM where the interference in phase space
makes the revival-collapse phenomenon in the atomic
inversion  more pronounced
when the revival time equals the half of that for the standard JCM with
initial coherent light \cite{buz6}.
The reason is that for the latter case the distribution of the
spectral components of the atomic inversion is one dimensional
distribution since only one
mode is involved, whereas here we have two-dimensional
superimposed distributions.
Further, we have obtained that the oscillations in $P(n)$ increase
if either $g$ or $t$
or both are increased. This fact is  clear if we look at the
problem as evolution of cat states under the action of two-mode
squeeze operator, where the squeezing parameter in this case is $r=gt$
and  the oscillations become more pronounced for  the
large values of $r$ \cite{caves,{ba4}}.
In conclusion, for the system under consideration the origin
of the oscillations
in the sum photon-number distribution  is two-fold: (i) The interference
in phase space. (ii) The strong coupling between the signal and idler modes
of the system in the course of the interaction time.

On the other hand, we have found  by an explicit algebraic
calculation for all quantities
studied in this paper excepting quadrature squeezing  that
the contributions associated with  elements of $\hat{\rho}_{\rm SI}(0)$ and
$\hat{\rho}_{\rm AI}(0)$
 involve  $\cos (\phi_{1}\pm\phi_{2})$ and
$\cos \phi_{1,2}$, respectively.  This fact  together with the information
included in Figs. 7 show that $P(n)$ can collapse
and yield decoherence in relation to distribution for statistical
mixture part using specific types of cat states initially.
More illustratively,  preparing one of the modes initially in ECS (or OCS)
and the other in YSS or vice versa, the $P(n)$ evolves as described by
Fig. 7a (for the same values of the parameters).
This shows  how one can  decohere the system
apart from the amplification nature of the system and
without coupling it to the environment.
We call such  a type of decoherence a phase decoherence.
Actually, for some quantities the contribution of off-diagonal elements
of the density matrix is responsible for  the nonclassical effects,
e.g. as we will see below in the
compound-mode reduced factorial moments.
Then we can make a good estimation of these phenomena using such
a property.
%%%%%%%%%%%%%%%%%%%%%%%%%%%%%%%%%%%%%%%%%%%%%%%%%%%%%%%%%%%%%%%%%%%%%%%
\begin{figure}[h]%
 \centering
 \subfigure[]{ \includegraphics[width=8cm]{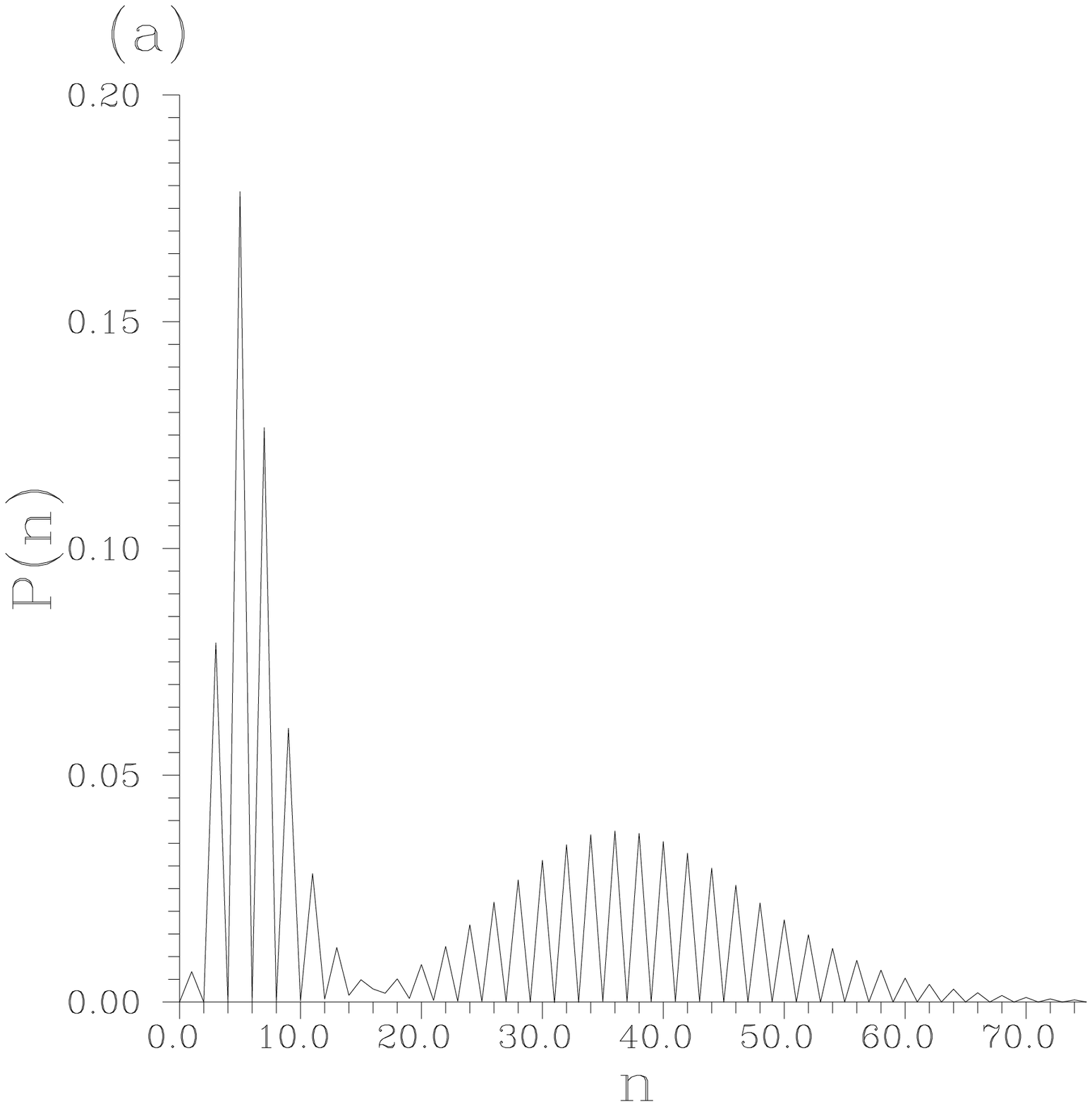}}
\subfigure[]{ \includegraphics[width=8cm]{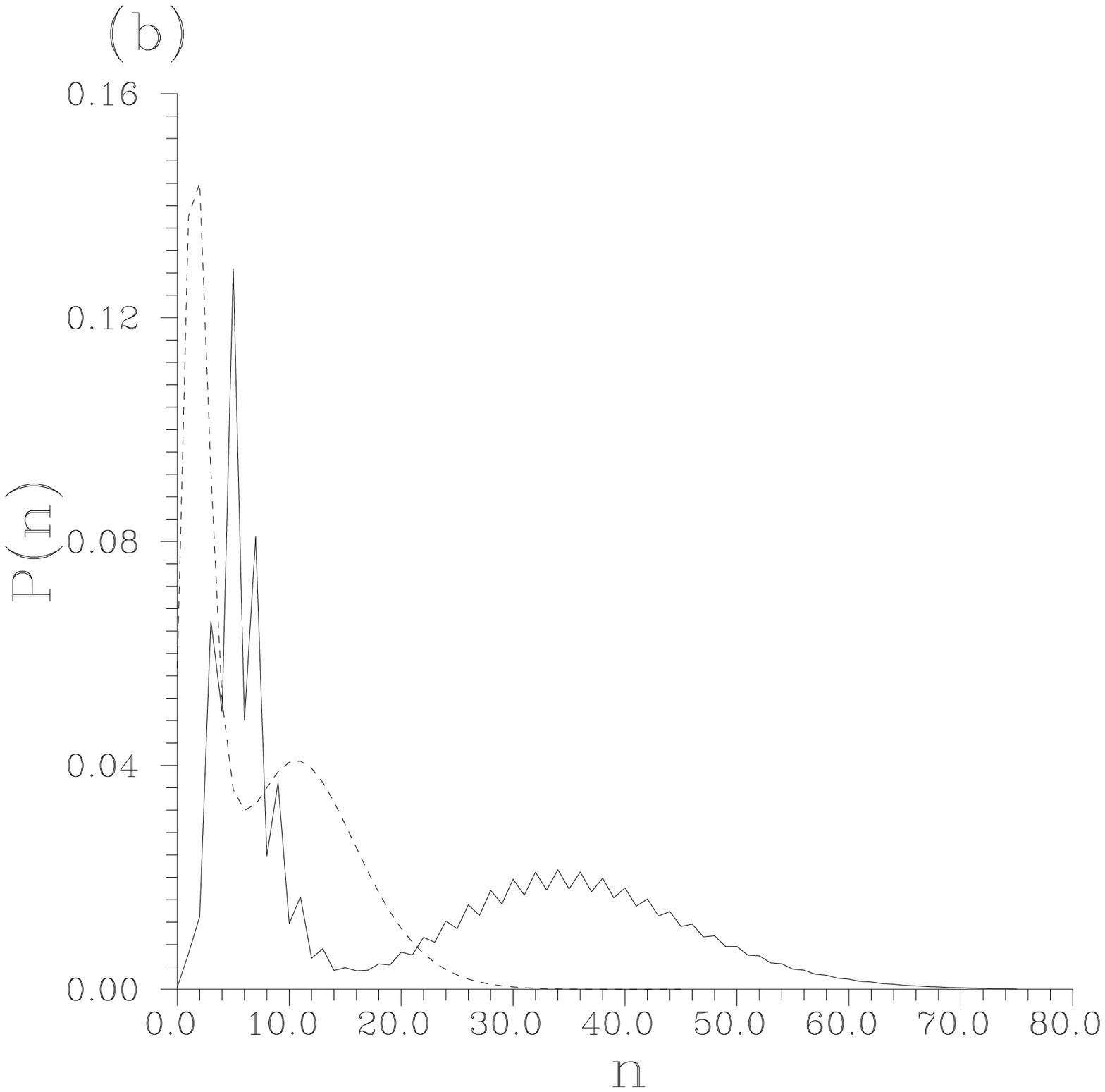}}
\caption{Sum photon-number distribution  for a) undamped case with
the values of the parameters as those of the solid curve in Fig.
3;
 b) damped case
with $(g, \bar{n},\gamma)=(0.5,0.5,2g-0.9)$ (underdamped case-solid curve)
and  $(0.5,0.5,2g+0.1)$ (overdamped case-dashed curve).}
\end{figure}
%%%%%%%%%%%%%%%%%%%%%%%%%%%%%%%%%%%%%%%%%%%%%%%%%%%%%%%%%%%%%%%
The final remark is that the nonclassical effects for the compound-mode case
are much richer than those for the single-mode case.
 This results from the strong quantum correlation
between the signal and idler
modes, which manifests itself as the summation in  expressions
(\ref{18})  and (\ref{19}).
To make this point  clear, we give Fig. 8a for the $P(n)$
and for the same situation as that of the solid curve in Fig. 3.
The comparison between these two figures is instructive.
Further, Fig. 8b includes information on the damped case
where one can observe smooth oscillations for the underdamped case
(solid curve) and two-peak structure for the overdamped case (dashed curve).
Specifically, for the overdamped case the off-diagonal elements
are completely suppressed.

%%%%%%%%%%%%%%%%%%%%%%%%%%%%%%%%%%%%%%%%%%%%%%%%%%%%%%%%%%%%%%
\subsection{Two-mode squeezing and reduced factorial moments}
%%%%%%%%%%%%%%%%%%%%%%%%%%%%%%%%%%%%%%%%%%%%%%%%%%%%%%%%%%%%%%
Firstly, we will discuss two-mode squeezing, which can be measured by
heterodyne detection where squeezing is carried jointly by two modes of
different frequencies and the local oscillator has a frequency midway
in between \cite{lodoun}.
We proceed taking into account that the nondegenerate parametric
amplifier is well described by the
two-mode squeeze operator.
Actually, two-mode squeezing phenomenon is essentially connected with the
states upon which such operator acts. In other words, squeezing may
not exist for
some specific states even if they include two-mode squeeze operator in
their structures.
To show this  we give the  form of the two-mode squeezing factor $Q$
 when the signal and idler modes  are initially prepared in
cat states, excluding losses,
 considering $\alpha_{j}, j=1,2$, are real and $\phi=0$. In this case the
squeezing factor $Q$ takes the form
\begin{equation}
Q(t)
=\frac{1}{2}[Q_{1}(t)+Q_{2}(t)],
\label{29}
\end{equation}
where $Q_{j}(t), j=1,2$, represent the corresponding single-mode  squeezing
factors  of the
 signal and idler modes, respectively.
From expression (\ref{29}) it is clear that
 the correlation between the signal and idler modes does not exist
since $\langle \hat{Y}_{1} \hat{Y}_{2}\rangle=
\langle \hat{Y}_{1}\rangle \langle\hat{Y}_{2}\rangle$, where
$\langle\hat{Y}_{j}\rangle, j=1,2$, are the expectation values of the
$Y$-quadrature  of the signal
 and idler modes.
The existence of correlation between these
two modes is important to obtain squeezing in the compound modes
even if the individual modes are not themselves squeezed \cite{ba4}.
Expression (\ref{29}) shows that interference in phase space can
destroy squeezing and also that the rates of degradation of the
two-mode and single-mode
squeezing are on the same level for symmetrical losses.
Also it is obvious that  the two-mode squeezing can be
realized if at least one of the two modes (signal or idler) can exhibit
single-mode  squeezing. This is necessary but not sufficient condition.
Further, the maximum squeezing may be produced when
both the signal and idler modes  exhibit maximum single-mode squeezing.
This reflects the importance of the choice of the type of the initial
cat states.
From the discussion in section 3 it is obvious that when the two modes
are initially in OCS, the interaction cannot generate two-mode squeezing.
%%%%%%%%%%%%%%%%%%%%%%%%%%%%%%%%%%%%%%%%%%%%%%%%%%%%%%%%%%%%%%%%%%%%%%%%%%%
\begin{figure}[h]%
 \centering
 \subfigure[]{ \includegraphics[width=8cm]{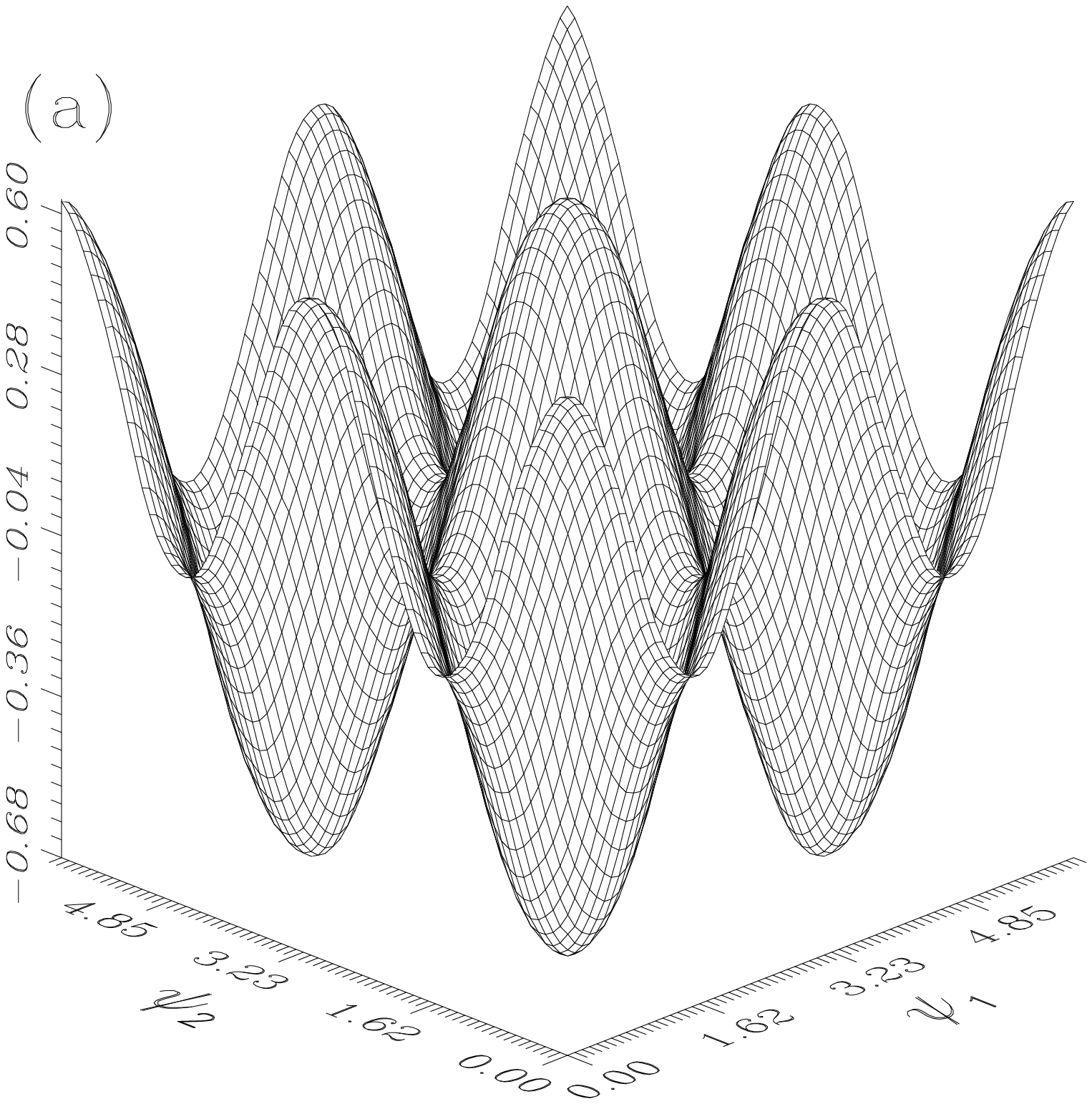}}
\subfigure[]{ \includegraphics[width=8cm]{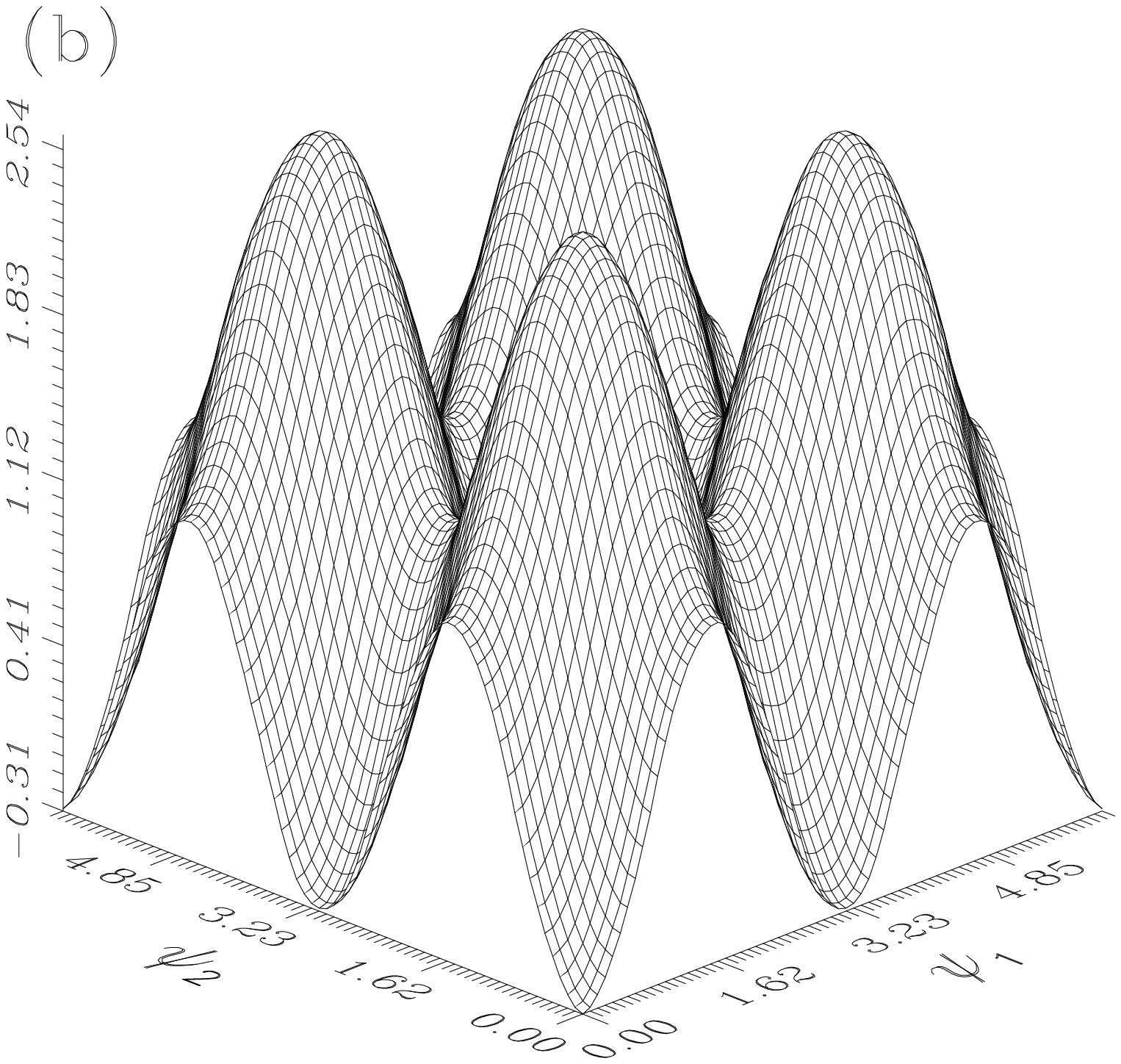}} \caption{The
compound two-mode squeezing against $\psi_{j},j=1,2$, for undamped
case when the signal and idler modes are  initially prepared in
ECS for $t=0.2,|\alpha_{j} |=0.7, g=1, \phi=\frac{\pi}{2}$: a)
squeezing factor $S$; b) squeezing factor $Q$.}
\end{figure}
%%%%%%%%%%%%%%%%%%%%%%%%%%%%%%%%%%%%%%%%%%%%%%%%%%%%%%%%%%%%%%%%%%%%%%%%%%%%%

We now discuss the case that one of the modes is initially squeezed (ECS)
and the other is unsqueezed (OCS) for the same situation as that in (\ref{29}).
The required time to obtain squeezing in the compound-mode case  ($Q(t)<0$) is

\begin{equation}
\sinh (gt)<\sqrt{\frac{-|\alpha_{1}|^{2}(\tanh |\alpha_{1}|^{2}-1)
-|\alpha_{2}|^{2}(\coth |\alpha_{2}|^{2}-1)}{2(1+|\alpha_{1}|^{2}
\tanh |\alpha_{1}|^{2}+|\alpha_{2}|^{2}\coth |\alpha_{2}|^{2})}}.
\label{29a}
\end{equation}
This inequality gives relation between the squeeze time and the
``distances" between the states, which are forming the cats.
So that the length of the device can be adjusted to obtain squeezed light.
An analysis to the right-hand side of (\ref{29a}) gives
that the maximum scaled time (effective time) for squeezing is $\tau=gt\simeq 0.1769$ at
$(|\alpha_{1}|,|\alpha_{2}|)=(0.6,2.4)$.
It is worth mentioning at this point that the $Y$-component quadrature squeezing
of ECS takes on its maximum, whereas that of OCS vanishes.

%%%%%%%%%%%%%%%%%%%%%%%%%%%%%%%%%%%%%%%%%%%%%%%%%%%%%%%%%%%%%%%%%%%%%%%%%
\begin{figure}[h]%
 \centering
  \includegraphics[width=10cm]{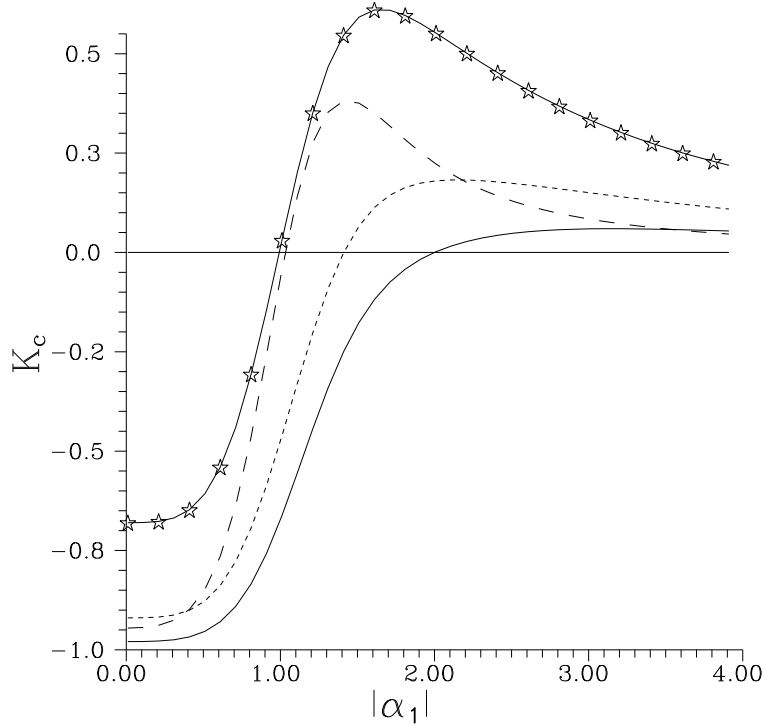}
\caption{The compound-mode reduced factorial moments
$(K_{c}=\langle W^{k}\rangle/\langle W\rangle^{k}-1)$ against
$|\alpha_{1} |$ for  $k=5, \psi=\pi/2, t=0.2,|\alpha_{2} |=0.25,
g=0.5$, and for the undamped case when the signal and idler modes
are  prepared initially in (OCS,OCS) (solid curve) and (OCS,ECS)
(long-dashed curve). The damped case is considered  for the same
situation as the solid curve, but for the underdamped case
(short-dashed curve) $(\gamma, \bar{n})=(2g-0.6, 0.5)$
 for the overdamped case (star-centred curve)
$(\gamma, \bar{n})=(2g+0.1, 0.5)$.
The straightline shows the antibunching bound.
}
\end{figure}
%%%%%%%%%%%%%%%%%%%%%%%%%%%%%%%%%%%%%%%%%%%%%%%%%%%%%%%%%%%%%%%%%%%%%%%%%%%

On the other hand, when the  phase information is included
(i.e. $\psi_{j}\neq 0, \phi\neq 0$)  the situation is improved
 in such a way that squeezing can be detected
in both quadratures  provided that
one of the modes is initially prepared in a squeezed-cat state, i.e.
in ECS or YSS.
The role of the phase is analyzed in
 Figs. 9a and b where we have plotted the squeezing factors corresponding to
$X$-component and $Y$-component, respectively, when the signal and idler
modes are initially prepared in ECS.
From these figures we can see that maximum squeezing occurs in $X$-component
and locates at
$(\psi_{1},\psi_{2})=(m_{1}\pi,m_{2}\pi), m_{j}=1/2,3/2$.
The opposite situation can be observed for the $Y$-component where
squeezing also exists but it is less pronounced  and locates at
$(\psi_{1},\psi_{2})=(m_{1}\pi,m_{2}\pi), m_{j}=0,1,2$,
i.e. the extreme values are exchanged.
Such a behaviour in the two quadratures indicates that the
uncertainty relation holds.   Further, for $\phi=-\pi/2$
 we obtained similar behaviour, however, the maximum squeezing values
 would be available in the $Y$-component.
Further similar behaviour can be seen if  the other
types of cat states  are used.
It is remarkable that the behaviour of squeezing in the present interaction
is quite different from that for both ECS \cite{yur1} and
 two-mode squeezed coherent states
\cite{ba4} where for both squeezing  exists only in one of the two
quadrature components;
 perfect squeezing (i.e. $100\%$ squeezing) occurs only for the two-mode
squeezed coherent states.
In conclusion
the squeezing phenomenon discussed here is a manifestation of co-operative
effects among the phases of the different components of the system.
Furthermore, by controlling  the phases in this device, squeezing can be
amplified and switched between the two quadratures.

We would like to conclude this subsection by
shedding  the light on the behaviour of the reduced factorial
moments for the compound-mode case.
In this case  the antibunched light can be measured
 if both the signal and idler modes are simultaneously detected
 by means of two photodetectors and
then  their outputs are correlated.
Generally, reduced factorial moments  can evolve to produce
nonclassical negative values  based on the types of initial cat states.
Further, we have
checked that the contribution of $\hat{\rho}_{\rm M}(0)$
 cannot  produce  antibunched light independently. These two facts
reflect the role
of the phase decoherence  property discussed above.
On the other hand, similar argument as that for the single-mode
case can be given here except
the case  when the two modes are in
YSS where a slight antibunching can be observed,
but only for certain  values of the
interaction parameters.
Fig. 10 shows the reduced factorial moments of the
compound-mode case for the given values of the parameters.
In this figure we have used the normalized factor
$K_{c} (=\langle W^{k}\rangle/\langle W\rangle^{k}-1)$,
where the nonclassical
effect occurs when $K_{c}<0$.
From this figure, it is interesting to observe that the behaviour
of the reduced
factorial moments when the (signal, idler) modes are in (OCS,OCS) is
very close to the behaviour  of the
second-order correlation function of OCS \cite{buz1} provided that
 $|\alpha_{j}|, j=1,2$, are finite.

\section{Conclusions}
In this paper we have analyzed the properties of the  dissipative
parametric amplifier when the signal and idler modes are initially
prepared
in Schr\"{o}dinger-cat states.  Needless to say  the present interaction
as described by the Hamiltonian (\ref{3})
and the density matrix (\ref{8}) is much more complicated
than that in the simple case of  a harmonic oscillator coupled with
the heat bath \cite{buz2}--\cite{filip}. After obtaining the solution of the
Heisenberg--Langevin equations quantum statistics of interacting modes have
been investigated based on the normally ordered characteristic functions.
The system of damping oscillators has been assumed to have a flat spectrum
and a chaotic distribution.

In general, there are two operations controlling the behaviour of the
interaction which are  the interference in phase space and the entanglement.
Intuitively, the initial  macroscopic cat states cannot be preserved in the
system.
Furthermore, for long-time interaction the initial nonclassical
effects of the cats
are degraded by the amplification dynamics inherent in the system and
the cumulative effects of dissipation.
Excluding dissipation we have shown that if the input to the device are
 squeezed (sub-Poissonian) cat states, then the output may be squeezed
 (antibunched) too
provided that the  interaction time and gain are finite.
Furthermore, the device could be used to amplify squeezing in
the compound-mode case.
On the contrary, the well-known role of the parametric amplifier
 as a source of perfect
squeezing in the compound-mode case rather than in single-mode case
fades out
for input cat states.
The photon-number distribution can exhibit oscillatory behaviour,
which is  more pronounced in the compound-mode case than in
the single-mode case, regardless of the types of
the initial cat states.  Also  we have shown that the
decoherence can arise in the system from the  decay
of the pump and the phase control.

For dissipation case we have considered in detail only two cases
which are related to underdamped  and overdamped regimes.
In these cases the interaction tends to eliminate the off-diagonal
elements of the density matrix and to affect the
diagonal elements.
It has been found that the dynamical subsystems (signal or
idler modes)  collapse to
the statistical mixture state or thermal state according to whether
the underdamped regime
or overdamping regime is considered.
Further, we have  also shown that the single-mode squeezing
at nonzero temperature of the
environment decreases much faster than that at zero temperature, however,
this situation is generally valid for all  quantities studied in this
paper.
In conclusion,  the losses uniformly distributed over the device
deteriorate its operational characteristics.
Of course, such deterioration is more pronounced  for the overdamped case.
Finally, the inclusion of  lossy mechanics in the system is of
a great interest
for accurate measurements.

%%%%%%%%%%%%%%%%%%%%%%%%%%%%%%%%%%%%%%%%%%%%%%%%%%%%%%%%%%%%%
\section*{ Acknowledgment}
%%%%%%%%%%%%%%%%%%%%%%%%%%%%%%%%%%%%%%%%%%%%%%%%%%%%%%%%%%%%%%

J. P., V. P. and F. A. A. E-O  acknowledge the partial support from the Project
LN00A015 of the Czech Ministry of Education.
One of us (M. S. A.) is grateful for the financial support
from the Project Math 1418/19 of the Research Centre, College of
Science, King Saud University.

%%%%%%%%%%%%%%%%%%%%%%%%%%%%%%%%%%%%%%%%%%%%%%%%%%%%%%%%%%%%%%%%%%
\begin{center}
{\bf Appendix A}
\end{center}
%%%%%%%%%%%%%%%%%%%%%%%%%%%%%%%%%%%%%%%%%%%%%%%%%%%%%%%%%%%%%%%%%%%%
In this appendix we give the explicit forms for the
time-dependent coefficients ($f_{j}(t),\Gamma_{jl}(t),
\Gamma^{'}_{jl}(t)$) of the solution
of the Heisenberg-Langevin equations
of the Hamiltonian (\ref{3}) (for details see \cite{in9,{in10}}):

${\displaystyle f_{1}(t)=\frac{1}{\sqrt{\epsilon}}
\exp\left[-\frac{(\gamma_{1}+\gamma_{2})t}{4}\right]
\left[
\sqrt{\epsilon}\cosh \left(\frac{\sqrt{\epsilon} }{4}t\right)
+(\gamma_{2}-\gamma_{1})\sinh \left(\frac{\sqrt{\epsilon} }{4}t\right)\right]
,}
\hfill (A.1)$

$\hfill$

${\displaystyle f_{2}(t)=\frac{4ig\exp{(i\phi)}}{\sqrt{\epsilon}}
\exp\left[-\frac{(\gamma_{1}+\gamma_{2})t}{4}\right]
\sinh \left(\frac{\sqrt{\epsilon} }{4}t\right)
,}
\hfill (A.2)$

$\hfill$

${\displaystyle f_{3}(t)=\frac{1}{\sqrt{\epsilon}}
\exp\left[-\frac{(\gamma_{1}+\gamma_{2})t}{4}\right]
\left[\sqrt{\epsilon}\cosh \left(\frac{\sqrt{\epsilon} }{4}t\right)
+(\gamma_{1}-\gamma_{2})\sinh \left(\frac{\sqrt{\epsilon} }{4}t\right)\right]
,}
\hfill (A.3)$

$\hfill$

${\displaystyle \Gamma_{jl}(t)=\frac{-ik_{jl}}{\epsilon^{'2}_{jl}-
\frac{\epsilon}{16}}
\exp\left[-\frac{(\gamma_{1}+\gamma_{2})t}{4}\right]
\left\{
\sqrt{\epsilon}
\left[\epsilon^{'}_{jl}+(-1)^{j+1}\frac{(\gamma_{2}-\gamma_{1})}{4}\right]
\right.
}
\hfill $

$\hfill$

${\displaystyle
\left.
\times\left[\exp(\epsilon^{'}_{jl}t)-
\cosh \left(\frac{\sqrt{\epsilon} }{4}t\right)\right]
 -
\left[\frac{\epsilon}{4}+(-1)^{j+1}(\gamma_{2}-\gamma_{1})\epsilon^{'}_{jl}\right]
\sinh \left(\frac{\sqrt{\epsilon} }{4}t\right)\right\}
,}  \quad j=1,2,
\hfill (A.4)$

$\hfill$

${\displaystyle \Gamma^{'}_{jl}(t)=\frac{-2ig \exp(i\phi) k_{jl}}{\epsilon^{'2}_{jl}-
\frac{\epsilon}{16}}
\exp\left[-\frac{(\gamma_{1}+\gamma_{2})t}{4}\right]
\left\{\frac{\sqrt{\epsilon}}{2}
\left[\exp(\epsilon^{'}_{jl}t)-
\cosh \left(\frac{\sqrt{\epsilon} }{4}t\right)\right]\right.
}\hfill $

$\hfill$

${\displaystyle
 -\left. 2\epsilon^{'}_{jl}
\sinh \left(\frac{\sqrt{\epsilon} }{4}t\right)\right\}
,}\quad j=1,2,
\hfill (A.5)$

\noindent where

${\displaystyle
\epsilon= (\gamma_{1}-\gamma_{2})^{2}+16g^{2}, \quad
\epsilon^{'}_{jl}=\frac{1}{4}(\gamma_{1}+\gamma_{2})+i(\omega_{j}-\varphi_{jl})
.} \hfill (A.6)$

%%%%%%%%%%%%%%%%%%%%%%%%%%%%%%%%%%%%%%%%%%%%%%%%%%%%%%%%%%%%%%
\begin{center}
{\bf Appendix B}
\end{center}
%%%%%%%%%%%%%%%%%%%%%%%%%%%%%%%%%%%%%%%%%%%%%%%%%%%%%%%%%%%%
In this appendix we write down  the explicit forms for the
 quantities $(B_{j{\cal N}}(t), D(t))$  in the expression (\ref{12}) (for details see \cite{in9,{in10}}):

${\displaystyle
B_{1{\cal N}}(t) =
B_{1{\cal N}}(\gamma_{1},\gamma_{2},\langle n_{1d}\rangle,\langle
n_{2d}\rangle,t)
=\frac{1}{\epsilon}\Bigl\{ 8g^{2}E_{1}
}
\hfill$

${\displaystyle
+\frac{\gamma_{1}\langle n_{1d}\rangle }
{\gamma_{1}\gamma_{2}-4g^{2}}\left[\left(\gamma_{2}\epsilon-4g^{2}(\gamma_{1}+\gamma_{2})
\right)E
-\sqrt{\epsilon}\left(\gamma_{2}(\gamma_{2}-\gamma_{1})+4g^{2}\right)F\right]
}
\hfill$

${\displaystyle
+\frac{4\gamma_{2}g^{2}(1+\langle n_{2d}\rangle) }
{\gamma_{1}\gamma_{2}-4g^{2}}[ (\gamma_{1}+\gamma_{2})E-\sqrt{\epsilon}F]
}
\hfill $

${\displaystyle
- 16g^{2}G[\gamma_{2}(1+\langle n_{2d}\rangle) -\gamma_{1}\langle
n_{1d}\rangle]\Bigr\}
,}
\hfill (B.1)$

$\hfill$

${\displaystyle
B_{2{\cal N}}(t) =B_{1{\cal N}}(\gamma_{2},\gamma_{1},\langle n_{2d}\rangle,
\langle n_{1d}\rangle,t) ,}
\hfill (B.2)$

$\hfill$

${\displaystyle
D(t)=\frac{2g\exp(-i\phi)}{\epsilon}\Bigl\{
(\gamma_{2}-\gamma_{1})E_{1}- \sqrt{\epsilon}F
}
\hfill$

${\displaystyle
+\frac{\gamma_{1}\langle n_{1d}\rangle }
{\gamma_{1}\gamma_{2}-4g^{2}}[(\gamma_{1}\gamma_{2}-\gamma_{2}^{2}-8g^{2})E
+\gamma_{2} \sqrt{\epsilon}F]
}
\hfill$

${\displaystyle
+\frac{\gamma_{2}(1+\langle n_{2d}\rangle) }
{\gamma_{1}\gamma_{2}-4g^{2}}[(\gamma_{1}\gamma_{2}-\gamma_{1}^{2}-8g^{2})E
+\gamma_{1} \sqrt{\epsilon}F]
}\hfill $

${\displaystyle
+2G[\gamma_{1}\langle n_{1d}\rangle (\gamma_{2}-\gamma_{1})
 +\gamma_{2}(1+\langle n_{2d}\rangle) (\gamma_{1}-\gamma_{2}) ]\Bigr\}
,}
\hfill (B.3)$

$\hfill$

${\displaystyle
\bar{\alpha}_{1}(t)={\alpha}^{*}_{1}f_{1}(t)+{\alpha}'_{2}f_{2}^{*}(t),
\quad \quad
\bar{\alpha}_{2}(t)={\alpha}'_{1}f_{2}^{*}(t)+{\alpha}^{*}_{2}f_{3}(t),}
\hfill$

${\displaystyle
\bar{\alpha}'_{1}(t)={\alpha}'_{1}f_{1}(t)+{\alpha}^{*}_{2}f_{2}(t),
\quad \quad
\bar{\alpha}'_{2}(t)={\alpha}^{*}_{1}f_{2}(t)+{\alpha}'_{2}f_{3}(t),}
\hfill (B.4)$

\noindent while

${\displaystyle
E=1-
\exp\left[-\frac{(\gamma_{1}+\gamma_{2})t}{2}\right]
\cosh \left(\frac{\sqrt{\epsilon} }{2}t\right)
,}
\hfill (B.5a)$

$\hfill$

${\displaystyle
E_{1}=
\exp\left[-\frac{(\gamma_{1}+\gamma_{2})t}{2}\right]
\left[\cosh \left(\frac{\sqrt{\epsilon} }{2}t\right)-1\right]
,}
\hfill (B.5b)$

$\hfill$

${\displaystyle
F=\exp\left[-\frac{(\gamma_{1}+\gamma_{2})t}{2}\right]
\sinh \left(\frac{\sqrt{\epsilon} }{2}t\right)
,}
\hfill (B.5c)$

$\hfill$

${\displaystyle
G=\frac{1}{\gamma_{1}+\gamma_{2}}\Bigl\{1-\exp\left[-\frac{(\gamma_{1}
+\gamma_{2})t}{2}\right]
\Bigr\}.}
\hfill (B.5d)$

A generalization to the case when the total density operator is
not factorized and the correlation between the system and
reservoirs is present, may be excluded \cite{collett}.

\end{document}